\input harvmac

\input amssym
\input epsf


\newfam\frakfam
\font\teneufm=eufm10
\font\seveneufm=eufm7
\font\fiveeufm=eufm5
\textfont\frakfam=\teneufm
\scriptfont\frakfam=\seveneufm
\scriptscriptfont\frakfam=\fiveeufm


\def\bb{
\font\tenmsb=msbm10
\font\sevenmsb=msbm7
\font\fivemsb=msbm5
\textfont1=\tenmsb
\scriptfont1=\sevenmsb
\scriptscriptfont1=\fivemsb
}



\newfam\dsromfam
\font\tendsrom=dsrom10
\textfont\dsromfam=\tendsrom
\def\ds{\fam\dsromfam \tendsrom}


\newfam\mbffam
\font\tenmbf=cmmib10
\font\sevenmbf=cmmib7
\font\fivembf=cmmib5
\textfont\mbffam=\tenmbf
\scriptfont\mbffam=\sevenmbf
\scriptscriptfont\mbffam=\fivembf


\newfam\mbfcalfam
\font\tenmbfcal=cmbsy10
\font\sevenmbfcal=cmbsy7
\font\fivembfcal=cmbsy5
\textfont\mbfcalfam=\tenmbfcal
\scriptfont\mbfcalfam=\sevenmbfcal
\scriptscriptfont\mbfcalfam=\fivembfcal


\newfam\mscrfam
\font\tenmscr=rsfs10
\font\sevenmscr=rsfs7
\font\fivemscr=rsfs5
\textfont\mscrfam=\tenmscr
\scriptfont\mscrfam=\sevenmscr
\scriptscriptfont\mscrfam=\fivemscr
\def\scr{\fam\mscrfam \tenmscr}




\def\tilde{\widetilde}
\def\t{\tilde}
\def\hat{\widehat}

\def\bar{\overline}
\def\b{\bar}
\def\bsq#1{{{\b{#1}}^{\lower 2.5pt\hbox{$\scriptstyle 2$}}}}
\def\bexp#1#2{{{\b{#1}}^{\lower 2.5pt\hbox{$\scriptstyle #2$}}}}
\def\dotexp#1#2{{{#1}^{\lower 2.5pt\hbox{$\scriptstyle #2$}}}}


\def\rt2{\sqrt{2}}
\def\half {{1 \over 2}}
\def\Re{\mathop{\rm Re}}
\def\Im{\mathop{\rm Im}}
\def\d{\partial}

\def\grad{\nabla}

\def\det{\mathop{\rm det}}


\font\tenbifull=cmmib10
\font\tenbimed=cmmib7
\font\tenbismall=cmmib5
\textfont9=\tenbifull \scriptfont9=\tenbimed
\scriptscriptfont9=\tenbismall

\mathchardef\bbGamma="7000
\mathchardef\bbDelta="7001
\mathchardef\bbPhi="7002
\mathchardef\bbAlpha="7003
\mathchardef\bbXi="7004
\mathchardef\bbPi="7005
\mathchardef\bbSigma="7006
\mathchardef\bbUpsilon="7007
\mathchardef\bbTheta="7008
\mathchardef\bbPsi="7009
\mathchardef\bbOmega="700A
\mathchardef\bbalpha="710B
\mathchardef\bbbeta="710C
\mathchardef\bbgamma="710D
\mathchardef\bbdelta="710E
\mathchardef\bbepsilon="710F
\mathchardef\bbzeta="7110
\mathchardef\bbeta="7111
\mathchardef\bbtheta="7112
\mathchardef\bbiota="7113
\mathchardef\bbkappa="7114
\mathchardef\bblambda="7115
\mathchardef\bbmu="7116
\mathchardef\bbnu="7117
\mathchardef\bbxi="7118
\mathchardef\bbpi="7119
\mathchardef\bbrho="711A
\mathchardef\bbsigma="711B
\mathchardef\bbtau="711C
\mathchardef\bbupsilon="711D
\mathchardef\bbphi="711E
\mathchardef\bbchi="711F
\mathchardef\bbpsi="7120
\mathchardef\bbomega="7121
\mathchardef\bbvarepsilon="7122
\mathchardef\bbvartheta="7123
\mathchardef\bbvarpi="7124
\mathchardef\bbvarrho="7125
\mathchardef\bbvarsigma="7126
\mathchardef\bbvarphi="7127


\def\alphadot{{\dot\alpha}}




\def\CA{{\cal A}}

\def\CD{{\cal D}}

\def\CH{{\cal H}}

\def\CJ{{\cal J}}
\def\CK{{\cal K}}
\def\CL{{\cal L}}
\def\CM{{\cal M}}
\def\CN{{\cal N}}
\def\CO{{\cal O}}

\def\CR{{\cal R}}
\def\CS{{\cal S}}

\def\CV{{\cal V}}


\def\1{{\ds 1}}
\def\R{\hbox{$\bb R$}}


\def\ep{\varepsilon}

\def\fz{{z } 
}
\def\fr{{ r} 
}
\def\fq{{ q} 
}

\noblackbox

\def\unit{\relax{\rm 1\kern-.26em I}}
\def\nada{\relax{\rm 0\kern-.30em l}}
\def\tilde{\widetilde}
\def\t{\tilde}
\def\alphadot{{\dot \alpha}}

\def\det{{\rm det}}

\noblackbox
\def\IL{\relax{\rm I\kern-.18em L}}
\def\IH{\relax{\rm I\kern-.18em H}}
\def\IR{\relax{\rm I\kern-.18em R}}
\def\IC{\relax\hbox{$\inbar\kern-.3em{\rm C}$}}
\def\IZ{\relax\ifmmode\mathchoice
{\hbox{\cmss Z\kern-.4em Z}}{\hbox{\cmss Z\kern-.4em Z}} {\lower.9pt\hbox{\cmsss Z\kern-.4em Z}}
{\lower1.2pt\hbox{\cmsss Z\kern-.4em Z}}\else{\cmss Z\kern-.4em Z}\fi}
\def\CM {{\cal M}}
\def\CN {{\cal N}}
\def\CR {{\cal R}}
\def\CD {{\cal D}}

\def\CJ {{\cal J}}
\def\partialslash{\not{\hbox{\kern-2pt $\partial$}}}

\def\CL {{\cal L}}
\def\CV {{\cal V}}
\def\CO {{\cal O}}

\def\CH {{\cal H}}

\def\CS {{\cal S}}
\def\CA{{\cal A}}
\def\CK{{\cal K}}
\def\CM {{\cal M}}
\def\CN {{\cal N}}

\def\CO {{\cal O}}

\def\CV{{\cal V }}

\def\CS {{\cal S }}

\def\det{{\rm det}}

\font\manual=manfnt \def\dbend{\lower3.5pt\hbox{\manual\char127}}

\def\IZ{\relax\ifmmode\mathchoice
{\hbox{\cmss Z\kern-.4em Z}}{\hbox{\cmss Z\kern-.4em Z}} {\lower.9pt\hbox{\cmsss Z\kern-.4em Z}}
{\lower1.2pt\hbox{\cmsss Z\kern-.4em Z}}\else{\cmss Z\kern-.4em Z}\fi}
\def\half {{1\over 2}}

\def\bar{\overline}
\def\CS{{\cal S}}
\def\CH{{\cal H}}

\def\rt2{\sqrt{2}}
\def\irt2{{1\over\sqrt{2}}}

\def\t{\tilde}
\def\hat{\widehat}
\def\slashchar#1{\setbox0=\hbox{$#1$}           
   \dimen0=\wd0                                 
   \setbox1=\hbox{/} \dimen1=\wd1               
   \ifdim\dimen0>\dimen1                        
      \rlap{\hbox to \dimen0{\hfil/\hfil}}      
      #1                                        
   \else                                        
      \rlap{\hbox to \dimen1{\hfil$#1$\hfil}}   
      /                                         
   \fi}

\def\foursqr#1#2{{\vcenter{\vbox{
    \hrule height.#2pt
    \hbox{\vrule width.#2pt height#1pt \kern#1pt
    \vrule width.#2pt}
    \hrule height.#2pt
    \hrule height.#2pt
    \hbox{\vrule width.#2pt height#1pt \kern#1pt
    \vrule width.#2pt}
    \hrule height.#2pt
        \hrule height.#2pt
    \hbox{\vrule width.#2pt height#1pt \kern#1pt
    \vrule width.#2pt}
    \hrule height.#2pt
        \hrule height.#2pt
    \hbox{\vrule width.#2pt height#1pt \kern#1pt
    \vrule width.#2pt}
    \hrule height.#2pt}}}}
\def\psqr#1#2{{\vcenter{\vbox{\hrule height.#2pt
    \hbox{\vrule width.#2pt height#1pt \kern#1pt
    \vrule width.#2pt}
    \hrule height.#2pt \hrule height.#2pt
    \hbox{\vrule width.#2pt height#1pt \kern#1pt
    \vrule width.#2pt}
    \hrule height.#2pt}}}}
\def\sqr#1#2{{\vcenter{\vbox{\hrule height.#2pt
    \hbox{\vrule width.#2pt height#1pt \kern#1pt
    \vrule width.#2pt}
    \hrule height.#2pt}}}}

\def\figin{\epsfcheck\figin}\def\figins{\epsfcheck\figins}
\def\epsfcheck{\ifx\epsfbox\UnDeFiNeD
\message{(NO epsf.tex, FIGURES WILL BE IGNORED)}
\gdef\figin##1{\vskip2in}\gdef\figins##1{\hskip.5in}
\else\message{(FIGURES WILL BE INCLUDED)}%
\gdef\figin##1{##1}\gdef\figins##1{##1}\fi}
\def\DefWarn#1{}
\def\figinsert{\goodbreak\midinsert}
\def\ifig#1#2#3{\DefWarn#1\xdef#1{fig.~\the\figno}
\writedef{#1\leftbracket fig.\noexpand~\the\figno}%
\figinsert\figin{\centerline{#3}}\medskip\centerline{\vbox{\baselineskip12pt \advance\hsize by
-1truein\noindent\footnotefont{\bf Fig.~\the\figno:\ } \it#2}}
\bigskip\endinsert\global\advance\figno by1}


\lref\FestucciaWS{
  G.~Festuccia and N.~Seiberg,
  ``Rigid Supersymmetric Theories in Curved Superspace,''
JHEP {\bf 1106}, 114 (2011).
[arXiv:1105.0689 [hep-th]].
}

\lref\BarnesBW{
  E.~Barnes, E.~Gorbatov, K.~A.~Intriligator and J.~Wright,
  ``Current correlators and AdS/CFT geometry,''
Nucl.\ Phys.\ B {\bf 732}, 89 (2006).
[hep-th/0507146].
}

\lref\VdBthesis{
  F.~van de Bult,
  ``Hyperbolic Hypergeometric Functions,''\vskip1pt
  [http://www.its.caltech.edu/~vdbult/Thesis.pdf].
}

\lref\ScherkZR{
  J.~Scherk and J.~H.~Schwarz,
  ``How to Get Masses from Extra Dimensions,''
Nucl.\ Phys.\ B {\bf 153}, 61 (1979).
}

\lref\Courant{
  R.~Courant and D.~Hilbert,
  ``Methods of Mathematical Physics, vol. II''
Interscience Publishers, (1962).
}

\lref\Nirenberg{
 L.~Nirenberg,
  ``Lectures on linear partial differential equations,''
Conference Board of the Mathematical Sciences, Regional Conference Series in Mathematics, No. 17, American Mathematical Society, (1973).
}

\lref\Newlander{
A.~Newlander and L.~Nirenberg,
  ``Complex analytic coordinates in almost complex manifolds,"
Ann. of Math. (2), {\bf 65}, 391, (1957).
}

\lref\Blair{
D.~E.~Blair, ``Riemannian Geometry of Contact and Symplectic Manifolds,''
Birkh\"auser, (2010).
}

\lref\Sasaki{
S.~Sasaki and Y.~Hatakeyama,
``On differentiable manifolds with certain structures which are closely related to almost contact structure. II,"
T\^ohoku Math.\ J.\ (2),
   {\bf 13}, 281, (1961).}

\lref\PestunRZ{
  V.~Pestun,
  ``Localization of gauge theory on a four-sphere and supersymmetric Wilson loops,''
Commun.\ Math.\ Phys.\  {\bf 313}, 71 (2012).
[arXiv:0712.2824 [hep-th]].
}

\lref\KapustinKZ{
  A.~Kapustin, B.~Willett and I.~Yaakov,
  ``Exact Results for Wilson Loops in Superconformal Chern-Simons Theories with Matter,''
JHEP {\bf 1003}, 089 (2010).
[arXiv:0909.4559 [hep-th]].
}

\lref\HamaAV{
  N.~Hama, K.~Hosomichi and S.~Lee,
  ``Notes on SUSY Gauge Theories on Three-Sphere,''
JHEP {\bf 1103}, 127 (2011).
[arXiv:1012.3512 [hep-th]].
}

\lref\JafferisUN{
  D.~L.~Jafferis,
  ``The Exact Superconformal R-Symmetry Extremizes Z,''
JHEP {\bf 1205}, 159 (2012).
[arXiv:1012.3210 [hep-th]].
}

\lref\HamaEA{
  N.~Hama, K.~Hosomichi and S.~Lee,
  ``SUSY Gauge Theories on Squashed Three-Spheres,''
JHEP {\bf 1105}, 014 (2011).
[arXiv:1102.4716 [hep-th]].
}

\lref\DolanRP{
  F.~A.~H.~Dolan, V.~P.~Spiridonov and G.~S.~Vartanov,
  ``From 4d superconformal indices to 3d partition functions,''
Phys.\ Lett.\ B {\bf 704}, 234 (2011).
[arXiv:1104.1787 [hep-th]].
}

\lref\GaddeIA{
  A.~Gadde and W.~Yan,
  ``Reducing the 4d Index to the $S^3$ Partition Function,''
[arXiv:1104.2592 [hep-th]].
}

\lref\ImamuraUW{
  Y.~Imamura,
  ``Relation between the 4d superconformal index and the $S^3$ partition function,''
JHEP {\bf 1109}, 133 (2011).
[arXiv:1104.4482 [hep-th]].
}

\lref\ImamuraWG{
  Y.~Imamura and D.~Yokoyama,
  ``N=2 supersymmetric theories on squashed three-sphere,''
Phys.\ Rev.\ D {\bf 85}, 025015 (2012).
[arXiv:1109.4734 [hep-th]].
}

\lref\MartelliFU{
  D.~Martelli, A.~Passias and J.~Sparks,
  ``The Gravity dual of supersymmetric gauge theories on a squashed three-sphere,''
Nucl.\ Phys.\ B {\bf 864}, 840 (2012).
[arXiv:1110.6400 [hep-th]].
}

\lref\MartelliFW{
  D.~Martelli and J.~Sparks,
  ``The gravity dual of supersymmetric gauge theories on a biaxially squashed three-sphere,''
Nucl.\ Phys.\ B {\bf 866}, 72 (2013).
[arXiv:1111.6930 [hep-th]].
}

\lref\HamaBG{
  N.~Hama and K.~Hosomichi,
  ``Seiberg-Witten Theories on Ellipsoids,''
Erratum-ibid.\  {\bf 1210}, 051 (2012).
[arXiv:1206.6359 [hep-th]].
}

\lref\BeniniUI{
  F.~Benini and S.~Cremonesi,
  ``Partition functions of N=(2,2) gauge theories on $S^2$ and vortices,''
[arXiv:1206.2356 [hep-th]].
}

\lref\DoroudXW{
  N.~Doroud, J.~Gomis, B.~Le Floch and S.~Lee,
  ``Exact Results in D=2 Supersymmetric Gauge Theories,''
[arXiv:1206.2606 [hep-th]].
}

\lref\GomisWY{
  J.~Gomis and S.~Lee,
  ``Exact Kahler Potential from Gauge Theory and Mirror Symmetry,''
[arXiv:1210.6022 [hep-th]].
}

\lref\KallenCS{
  J.~Kallen and M.~Zabzine,
  ``Twisted supersymmetric 5D Yang-Mills theory and contact geometry,''
JHEP {\bf 1205}, 125 (2012).
[arXiv:1202.1956 [hep-th]].
}

\lref\HosomichiEK{
  K.~Hosomichi, R.~-K.~Seong and S.~Terashima,
  ``Supersymmetric Gauge Theories on the Five-Sphere,''
Nucl.\ Phys.\ B {\bf 865}, 376 (2012).
[arXiv:1203.0371 [hep-th]].
}

\lref\JafferisIV{
  D.~L.~Jafferis and S.~S.~Pufu,
  ``Exact results for five-dimensional superconformal field theories with gravity duals,''
[arXiv:1207.4359 [hep-th]].
}

\lref\ImamuraXG{
  Y.~Imamura,
  ``Supersymmetric theories on squashed five-sphere,''
[arXiv:1209.0561 [hep-th]].
}

\lref\KallenNY{
  J.~Kallen,
 ``Cohomological localization of Chern-Simons theory,''
JHEP {\bf 1108}, 008 (2011).
[arXiv:1104.5353 [hep-th]].
}

\lref\SamtlebenGY{
  H.~Samtleben and D.~Tsimpis,
  ``Rigid supersymmetric theories in 4d Riemannian space,''
JHEP {\bf 1205}, 132 (2012).
[arXiv:1203.3420 [hep-th]].
}

\lref\OhtaEV{
  K.~Ohta and Y.~Yoshida,
  ``Non-Abelian Localization for Supersymmetric Yang-Mills-Chern-Simons Theories on Seifert Manifold,''
[arXiv:1205.0046 [hep-th]].
}

\lref\KlareGN{
  C.~Klare, A.~Tomasiello and A.~Zaffaroni,
  ``Supersymmetry on Curved Spaces and Holography,''
JHEP {\bf 1208}, 061 (2012).
[arXiv:1205.1062 [hep-th]].
}

\lref\DumitrescuHA{
  T.~T.~Dumitrescu, G.~Festuccia and N.~Seiberg,
  ``Exploring Curved Superspace,''
JHEP {\bf 1208}, 141 (2012).
[arXiv:1205.1115 [hep-th]].
}

\lref\DumitrescuAT{
  T.~T.~Dumitrescu and G.~Festuccia,
  ``Exploring Curved Superspace (II),''
[arXiv:1209.5408 [hep-th]].
}

\lref\BeemMB{
  C.~Beem, T.~Dimofte and S.~Pasquetti,
  ``Holomorphic Blocks in Three Dimensions,''
[arXiv:1211.1986 [hep-th]].
}

\lref\CassaniRI{
  D.~Cassani, C.~Klare, D.~Martelli, A.~Tomasiello and A.~Zaffaroni,
  ``Supersymmetry in Lorentzian Curved Spaces and Holography,''
[arXiv:1207.2181 [hep-th]].
}

\lref\LiuBI{
  J.~T.~Liu, L.~A.~Pando Zayas and D.~Reichmann,
  ``Rigid Supersymmetric Backgrounds of Minimal Off-Shell Supergravity,''
JHEP {\bf 1210}, 034 (2012).
[arXiv:1207.2785 [hep-th]].
}

\lref\JafferisZI{
  D.~L.~Jafferis, I.~R.~Klebanov, S.~S.~Pufu and B.~R.~Safdi,
  ``Towards the F-Theorem: N=2 Field Theories on the Three-Sphere,''
JHEP {\bf 1106}, 102 (2011).
[arXiv:1103.1181 [hep-th]].
}

\lref\ClossetVG{
  C.~Closset, T.~T.~Dumitrescu, G.~Festuccia, Z.~Komargodski and N.~Seiberg,
  ``Contact Terms, Unitarity, and F-Maximization in Three-Dimensional Superconformal Theories,''
JHEP {\bf 1210}, 053 (2012).
[arXiv:1205.4142 [hep-th]].
}

\lref\ClossetVP{
  C.~Closset, T.~T.~Dumitrescu, G.~Festuccia, Z.~Komargodski and N.~Seiberg,
  ``Comments on Chern-Simons Contact Terms in Three Dimensions,''
JHEP {\bf 1209}, 091 (2012).
[arXiv:1206.5218 [hep-th]].
}

\lref\DumitrescuIU{
  T.~T.~Dumitrescu and N.~Seiberg,
  ``Supercurrents and Brane Currents in Diverse Dimensions,''
JHEP {\bf 1107}, 095 (2011).
[arXiv:1106.0031 [hep-th]].
}

\lref\KomargodskiRB{
  Z.~Komargodski and N.~Seiberg,
  ``Comments on Supercurrent Multiplets, Supersymmetric Field Theories and Supergravity,''
JHEP {\bf 1007}, 017 (2010).
[arXiv:1002.2228 [hep-th]].
}

\lref\SohniusTP{
  M.~F.~Sohnius and P.~C.~West,
  ``An Alternative Minimal Off-Shell Version of N=1 Supergravity,''
Phys.\ Lett.\ B {\bf 105}, 353 (1981).
}

\lref\SohniusFW{
  M.~Sohnius and P.~C.~West,
  ``The Tensor Calculus And Matter Coupling Of The Alternative Minimal Auxiliary Field Formulation Of N=1 Supergravity,''
Nucl.\ Phys.\ B {\bf 198}, 493 (1982).
}

\lref\RocekBK{
  M.~Rocek and P.~van Nieuwenhuizen,
  ``N $\geq$ 2 supersymmetric Chern-Simons terms as d = 3 extended conformal supergravity,''
Class.\ Quant.\ Grav.\  {\bf 3}, 43 (1986).
}

\lref\AchucarroVZ{
  A.~Achucarro and P.~K.~Townsend,
  ``A Chern-Simons Action for Three-Dimensional anti-De Sitter Supergravity Theories,''
Phys.\ Lett.\ B {\bf 180}, 89 (1986).
}

\lref\IzquierdoJZ{
  J.~M.~Izquierdo and P.~K.~Townsend,
  ``Supersymmetric space-times in (2+1) adS supergravity models,''
Class.\ Quant.\ Grav.\  {\bf 12}, 895 (1995).
[gr-qc/9501018].
}

\lref\KuzenkoXG{
  S.~M.~Kuzenko, U.~Lindstrom and G.~Tartaglino-Mazzucchelli,
  ``Off-shell supergravity-matter couplings in three dimensions,''
JHEP {\bf 1103}, 120 (2011).
[arXiv:1101.4013 [hep-th]].
}

\lref\KuzenkoBC{
  S.~M.~Kuzenko, U.~Lindstrom and G.~Tartaglino-Mazzucchelli,
  ``Three-dimensional (p,q) AdS superspaces and matter couplings,''
JHEP {\bf 1208}, 024 (2012).
[arXiv:1205.4622 [hep-th]].
}

\lref\RocekWIP{
S.~M.~Kuzenko, U.~Lindstr\"om, M.~Rocek, I.~Sachs, G.~Tartaglino-Mazzucchelli, to appear.
}

\lref\Kosmann{
  Y.~Kosmann,
  ``D\'eriv\'ees de Lie des spineurs,''
Ann.\ di Matematica Pura e Appl.\  {\bf 91}, 317�395 (1972).
}

\lref\Isenberg{
  J.~Isenberg and V.~Moncrief
  ``On spacetimes containing Killing vector fields with non-closed orbits,''
Classical and Quantum Gravity, 1992, Jul 9, Volume 9, Issue 7, 1683-1691.
}

\lref\ImamuraSU{
  Y.~Imamura and S.~Yokoyama,
  ``Index for three dimensional superconformal field theories with general R-charge assignments,''
JHEP {\bf 1104}, 007 (2011).
[arXiv:1101.0557 [hep-th]].
}

\lref\BeniniMF{
  F.~Benini, C.~Closset and S.~Cremonesi,
  ``Comments on 3d Seiberg-like dualities,''
JHEP {\bf 1110}, 075 (2011).
[arXiv:1108.5373 [hep-th]].
}

\lref\HoweZM{
  P.~S.~Howe, J.~M.~Izquierdo, G.~Papadopoulos and P.~K.~Townsend,
  ``New supergravities with central charges and Killing spinors in (2+1)-dimensions,''
Nucl.\ Phys.\ B {\bf 467}, 183 (1996).
[hep-th/9505032].
}

\lref\KuzenkoQG{
  S.~M.~Kuzenko,
  ``Prepotentials for N=2 conformal supergravity in three dimensions,''
JHEP {\bf 1212}, 021 (2012).
[arXiv:1209.3894 [hep-th]].
}

\lref\KuzenkoRD{
  S.~M.~Kuzenko and G.~Tartaglino-Mazzucchelli,
  ``Three-dimensional N=2 (AdS) supergravity and associated supercurrents,''
JHEP {\bf 1112}, 052 (2011).
[arXiv:1109.0496 [hep-th]].
}

\lref\KehagiasFH{
  A.~Kehagias and J.~G.~Russo,
  ``Global Supersymmetry on Curved Spaces in Various Dimensions,''
[arXiv:1211.1367 [hep-th]].
}

\lref\ClossetVRA{
  C.~Closset, T.~T.~Dumitrescu, G.~Festuccia and Z.~Komargodski,
  ``The Geometry of Supersymmetric Partition Functions,''
[arXiv:1309.5876 [hep-th]].
}


\rightline{PUPT-2432}
\rightline{WIS/20/12-DEC-DPPA}
\vskip-15pt
\Title{
} {\vbox{\centerline{Supersymmetric Field Theories}
\centerline{on Three-Manifolds}}}
\vskip-20pt
\centerline{Cyril Closset,$^1$ Thomas T. Dumitrescu,$^{2}$ Guido Festuccia,$^3$ and Zohar Komargodski$^{1,3}$}
\vskip15pt
\centerline{ $^{1}$ {\it Weizmann Institute of Science, Rehovot
76100, Israel}}
 \centerline{$^{2}$ {\it Department of Physics, Princeton University, Princeton, NJ 08544, USA}}
  \centerline{$^{3}${\it
Institute for Advanced Study, Princeton, NJ 08540, USA}}

\vskip35pt

\noindent We construct supersymmetric field theories on Riemannian three-manifolds~$\CM$, focusing on~$\CN=2$ theories with a~$U(1)_R$ symmetry. Our approach is based on the rigid limit of new minimal supergravity in three dimensions, which couples to the flat-space supermultiplet containing the~$R$-current and the energy-momentum tensor. The field theory on~$\CM$ possesses a single supercharge if and only if~$\CM$ admits an almost contact metric structure that satisfies a certain integrability condition.  This may lead to global restrictions on~$\CM$, even though we can always construct one supercharge on any given patch. We also analyze the conditions for the presence of additional supercharges. In particular, two supercharges of opposite~$R$-charge exist on every Seifert manifold. We present general supersymmetric Lagrangians on~$\CM$ and discuss their flat-space limit, which can be analyzed using the~$R$-current supermultiplet.  As an application, we show how the flat-space two-point function of the energy-momentum
tensor in~$\CN = 2$ superconformal theories can be calculated using localization on a squashed sphere.

\vskip10pt

\Date{December 2012}

\newsec{Introduction}

In this paper, we construct supersymmetric quantum field theories on Riemannian three-manifolds~$\CM$, focusing on three-dimensional~$\CN=2$ theories with a~$U(1)_R$ symmetry. We can place any such theory on~$\CM$ by minimally coupling it to the metric, but the resulting theory generally does not possess any rigid supersymmetry. Under certain conditions it is possible to add non-minimal couplings, such that the deformed theory is invariant under one or several rigid supercharges, which are also suitably deformed. We will analyze the conditions for~$\CM$ to admit one or several rigid supercharges, and explain how to write supersymmetric Lagrangians on~$\CM$.

Following~\refs{\PestunRZ\KapustinKZ\JafferisUN-\HamaAV}, much work has focused on supersymmetric theories on round and squashed spheres~\refs{\ImamuraSU\HamaEA\DolanRP\GaddeIA\ImamuraUW\ImamuraWG\MartelliFU\MartelliFW-\HamaBG}.\foot{For related work in two and five dimensions, see~\refs{\BeniniUI\DoroudXW-\GomisWY} and~\refs{\KallenCS\HosomichiEK\JafferisIV-\ImamuraXG}, respectively. Lorentzian spacetimes with rigid supersymmetry were studied in~\refs{\CassaniRI,\LiuBI}.} Some more general geometries were recently considered in~\refs{\KallenNY\SamtlebenGY\OhtaEV\KlareGN\DumitrescuHA\DumitrescuAT-\BeemMB}. A systematic approach to this subject was initiated in~\FestucciaWS, using background supergravity. If an off-shell formulation of dynamical supergravity is available, we can couple it to the field theory of interest and take a rigid limit by appropriately sending the Planck mass to infinity. This decouples the fluctuations of the supergravity fields and allows us to freeze them in arbitrary background configurations.\foot{For an approach based on on-shell supergravity, see~\KehagiasFH.}

The supergravity multiplet generally contains the metric~$g_{\mu\nu}$, the gravitino~$\psi_{\mu\alpha}$, and various auxiliary fields. A crucial feature of the rigid limit is that we do not eliminate these auxiliary fields by imposing their equations of motion. If there are supergravity transformations that leave a given background invariant, they give rise to a rigid supercharge of the field theory on this background. This formalism has recently been used to classify supersymmetric backgrounds for~$\CN=1$ theories in four dimensions~\refs{\KlareGN\DumitrescuHA-\DumitrescuAT}. In three-dimensional~$\CN=2$ theories with a~$U(1)_R$ symmetry, it has led to a proof of the~$F$-maximization principle (originally conjectured in~\refs{\JafferisUN,\JafferisZI}), the discovery of a novel superconformal anomaly, and the definition and exact computation of several previously inaccessible observables~\refs{\ClossetVG,\ClossetVP}.

The rigid limit introduced in~\FestucciaWS\ allows a clean separation between the supergravity background fields and the dynamical matter fields:
\medskip
\item{1.)} The supergravity sector determines the allowed supersymmetric backgrounds and the corresponding supersymmetry algebras. A given supergravity transformation~$\delta$ gives rise to a rigid supercharge if and only if the background is such that both~$\psi_{\mu\alpha}$ and~$\delta \psi_{\mu\alpha}$ vanish. The rigid supersymmetry algebra immediately follows from the algebra of supergravity transformations. This part of the analysis does not depend on the details of the matter sector.

\item{2.)} The supersymmetry transformations of the matter fields and the corresponding supersymmetric matter Lagrangians do not depend on the details of a given supersymmetric background. They are obtained from the supergravity transformations and Lagrangians by taking the rigid limit. We can also take the rigid limit of purely gravitational terms in the Lagrangian, which reduce to pure numbers in a given background. Such terms correspond to contact terms in correlation functions of operators in the field theory and played a crucial role in~\refs{\ClossetVG,\ClossetVP}.
\medskip
\noindent The separation of the two sectors represents a significant simplification over approaches that consider specific manifolds with fixed metrics and derive the supersymmetry transformations and the corresponding Lagrangians on a case-by-case basis.

At the linearized level, a supersymmetric field theory couples to supergravity through its supercurrent multiplet, which contains the energy-momentum tensor and the supersymmetry current, as well as other operators. The structure of supercurrents in three-dimensional~$\CN=2$ theories was analyzed in~\DumitrescuIU. (They are closely related to supercurrents in four-dimensional~$\CN=1$ theories~\refs{\KomargodskiRB,\DumitrescuIU}.) Theories with a~$U(1)_R$ symmetry admit a supercurrent~$\CR_\mu$, whose bottom component is the conserved~$R$-current.

Different supercurrent multiplets give rise to different off-shell formulations of supergravity, which in turn lead to different rigid limits. These rigid limits need not give rise to the same set of supersymmetric manifolds~$\CM$, even if the original supergravity theories were equivalent on-shell. In this paper, we are interested in three-dimensional~$\CN=2$ theories with a~$U(1)_R$ symmetry, and hence we need the rigid limit of the three-dimensional supergravity that couples to the~$\CR$-multiplet. By analogy with the four-dimensional case~\refs{\SohniusTP,\SohniusFW}, we refer to this theory as new minimal supergravity. While it has recently been studied in superspace~\refs{\KuzenkoXG\KuzenkoRD\KuzenkoBC-\KuzenkoQG}, a fully non-linear component formulation of three-dimensional new minimal supergravity is not readily available. (However, see the early work~\refs{\RocekBK\AchucarroVZ\IzquierdoJZ-\HoweZM}, as well as~\RocekWIP.) Therefore, the corresponding rigid limit cannot be obtained as an immediate consequence of known results.

In this paper, we will nevertheless obtain most of the rigid limit without working out the full non-linear supergravity, by relying on input from linearized supergravity and the reduction of four-dimensional new minimal supergravity. In particular, we will obtain the equations that govern supersymmetric backgrounds, as well as the supersymmetry algebra, transformation laws, and Lagrangians for the matter fields. However, we will not derive the rigid limit of pure supergravity terms. (Such terms were analyzed in~\refs{\ClossetVG,\ClossetVP} and will make a brief appearance in section~8.)

At the linearized level, new minimal supergravity in three dimensions was recently studied in~\refs{\KuzenkoRD,\ClossetVG,\ClossetVP}. In addition to the metric~$g_{\mu\nu}$, the theory contains two gravitini~$\psi_{\mu\alpha}$ and~$\t \psi_{\mu\alpha}$,\foot{In three Euclidean dimensions, the Lorentz group is~$SU(2)$. Therefore, all spinors are complex and carry undotted indices. In Lorentzian signature, the gravitini~$\psi_{\mu\alpha}$ and~$\t \psi_{\mu\alpha}$ are related by complex conjugation, but here they are independent complex spinors. See appendix~A for a summary of our conventions.} two Abelian gauge fields~$A_\mu$ and~$C_\mu$, and a two-form gauge field~$B_{\mu\nu}$. We will often use the dual field strengths
\eqn\dfsdef{\eqalign{& V^\mu = -i \ep^{\mu\nu\rho} \d_\nu C_\rho~, \qquad \grad_\mu V^\mu = 0~, \cr
&  H = {i \over 2} \ep^{\mu\nu\rho} \d_\mu B_{\nu\rho}~.}}
All of these fields couple to operators in the~$\CR$-multiplet of the field theory. In particular, the gauge field~$A_\mu$ couples to the~$U(1)_R$ current, which leads to invariance under local~$R$-transformations, and the gauge field~$C_\mu$ couples to the current that gives rise to the central charge~$Z$ in the conventional flat-space~$\CN=2$ supersymmetry algebra,
\eqn\flatspacesusy{\eqalign{& \{Q_\alpha, \t Q_\beta\} = 2 \gamma^\mu_{\alpha\beta} P_\mu + 2 i \ep_{\alpha\beta} Z~,\cr
& \{Q_\alpha, Q_\beta\} = 0~, \qquad \{\t Q_\alpha, \t Q_\beta\} = 0~.}}
Here~$Q_\alpha$ and~$\t Q_\alpha$ have~$R$-charge~$-1$ and~$+1$, respectively.

In section~2 we review the three-dimensional~$\CR$-multiplet and its coupling to linearized new minimal supergravity. We study the variations of the gravitini at the linearized level and argue that they completely determine the corresponding variations in the full non-linear theory, up to terms that vanish in the rigid limit:
\eqn\gravivar{\eqalign{& \delta \psi_\mu = 2 \left(\grad_\mu - i A_\mu\right)\zeta + H \gamma_\mu \zeta + 2 i V_\mu \zeta + \ep_{\mu\nu\rho} V^\nu \gamma^\rho \zeta + \cdots~,\cr
& \delta \t \psi_\mu = 2 \left(\grad_\mu + i A_\mu\right) \t \zeta + H \gamma_\mu \t \zeta - 2 i V_\mu \t \zeta - \ep_{\mu\nu\rho} V^\nu \gamma^\rho \t \zeta + \cdots~.}}
Here~$\zeta$ and~$\t \zeta$ are spinors parameterizing the supergravity transformation. They carry~$R$-charge~$+1$ and~$-1$, respectively. The ellipses in~\gravivar\ denote terms that vanish when the gravitini are set to zero. In Euclidean signature, $\zeta$ and~$\t \zeta$ are independent complex spinors and we allow the background fields~$A_\mu, V_\mu, H$ to be complex. However, we will always take the metric~$g_{\mu\nu}$ to be real.  A given configuration of the supergravity background fields preserves rigid supersymmetry if and only if~$\psi_\mu, \t \psi_\mu$, and both variations in~\gravivar\ vanish for some choice of~$\zeta$ and~$\t \zeta$. Moreover, we can always consider variations of definite~$R$-charge. Therefore, a rigid supercharge~$\delta_\zeta$ of~$R$-charge~$-1$ corresponds to a solution~$\zeta$ of
\eqn\ksei{ \left(\grad_\mu - i A_\mu\right) \zeta =  -\half H \gamma_\mu\zeta - i V_\mu\zeta -\half \ep_{\mu\nu\rho}V^\nu \gamma^\rho\zeta~,}
while a rigid supercharge~$\delta_{\t \zeta}$ of~$R$-charge~$+1$ corresponds to a solution~$\t \zeta$ of
\eqn\kseii{\left(\grad_\mu + iA_\mu\right) \t\zeta =  -\half H \gamma_\mu\t\zeta + i V_\mu\t\zeta +\half\ep_{\mu\nu\rho}V^\nu \gamma^\rho\t\zeta~.}
These equations will allow us to determine which three-manifolds~$\CM$ admit rigid supersymmetry. We will refer to~\ksei\ and~\kseii\ as Killing spinor equations. Similar equations were previously obtained in~\refs{\HoweZM,\KlareGN}.

In section~3 we begin our analysis of Riemannian three-manifolds~$\CM$ that admit solutions of~\ksei\ or~\kseii. (This problem was also studied in~\KlareGN.) We analyze general properties of such solutions and use them to construct various useful spinor bilinears. In particular, we show that a solution~$\zeta$ of~\ksei\ naturally defines an almost contact metric structure (ACMS) on~$\CM$, and similarly for a solution~$\t \zeta$ of~\kseii. Such a structure constitutes an odd-dimensional analogue of an almost Hermitian structure. (The basic properties of ACMS's are reviewed in appendix~B.) Since the Killing spinor equations~\ksei\ and~\kseii\ are partial differential equations, they only admit solutions if the background fields~$g_{\mu\nu}, A_\mu, V_\mu, H$ satisfy certain integrability conditions. Moreover, there may be global obstructions. We would like to formulate necessary and sufficient conditions for the existence of one or several solutions.

In section~4 we show that~$\CM$ admits one solution~$\zeta$ of~\ksei\ if and only if the ACMS defined by~$\zeta$ satisfies a certain integrability condition. This integrability condition endows~$\CM$ with a three-dimensional analogue of an integrable complex structure.\foot{This structure has  been studied in the mathematical literature, where it is known as a transversely holomorphic foliation (see~\ClossetVRA\ for details and references). We are grateful to Maxim Kontsevich for pointing this out to us.} The existence of such a structure represents a global restriction on~$\CM$. (See appendix~B for additional details.) The supercharge corresponding to~$\zeta$ transforms as a scalar on~$\CM$.

 In section~5 we analyze manifolds admitting multiple solutions of~\ksei\ and~\kseii. Two solutions~$\zeta$ and~$\t \zeta$ of opposite~$R$-charge exist on any circle bundle over a Riemann surface, i.e. a Seifert manifold. In this case~$K^\mu \d_\mu  = \zeta \gamma^\mu \t \zeta \d_\mu = \d_\psi$ is a real Killing vector and the metric is given by
\eqn\seifmet{ds^2 = \Omega(z, \b z)^2 \left(d \psi + a(z, \b z) dz + \b a(z, \b z)d \b z\right)^2 + c(z, \b z)^2 dz d \b z~.}
We must distinguish two cases: If~$K^\mu$ points along the circle fibers, it has closed orbits. In this case~$z$ in~\seifmet\ is a holomorphic coordinate on the Riemann surface. However, if the orbits of~$K^\mu$ are not closed, then it must have a component along the Riemann surface. In this case the manifold must admit at least a~$U(1) \times U(1)$ isometry, one along the circle fibers and one along the base. All previously known backgrounds that preserve two supercharges of opposite~$R$-charge with real~$K^\mu$, such as  \refs{\HamaEA,\MartelliFU}, can be understood in this way. We also analyze solutions with four supercharges. In particular, we explain how the round~$S^3$ of~\refs{\KapustinKZ\JafferisUN-\HamaAV}, the~$S^2 \times S^1$ of~\ImamuraSU, and the squashed~$S^3$ of~\refs{\ImamuraUW,\ImamuraWG} arise in our formalism.

In section~6 we develop the necessary tools to write supersymmetric Lagrangians on manifolds that admit solutions to the Killing spinor equations. We discuss the general form of the rigid supersymmetry algebra and its multiplets. This enables us to construct Lagrangians for gauge and matter fields.

In section~7 we expand the supersymmetric background fields of section~4 and the supersymmetric Lagrangians of section~6 around flat space, where they can be analyzed using the~$\CR$-multiplet. We find that the first-order deformation of the Lagrangian around flat space is~$Q$-exact and comment on the relevance of this observation for the parameter dependence of supersymmetric partition functions.

In section 8 we demonstrate the utility of our formalism by relating the flat-space two-point function of the~$R$-current and the energy-momentum tensor in an~$\CN=2$ superconformal field theory (SCFT) to the partition function of the same theory on the squashed three-sphere of~\refs{\ImamuraUW,\ImamuraWG}. This enables us to compute these two-point functions exactly using localization. (The two-point functions of flavor currents can be extracted from the partition function on a round three-sphere~\ClossetVG.)

Our conventions are summarized in appendix~A. In appendix~B, we collect some basic facts about almost contact structures and analyze the integrability condition introduced in section~4. Appendix~C reviews the rigid limit of four-dimensional new minimal supergravity. In appendix~D, we derive the three-dimensional supersymmetry algebra and superfield transformation laws discussed in section~6 by a twisted dimensional reduction from four dimensions.

\newsec{The~$\CR$-Multiplet and New Minimal Supergravity}

In this section we review the~$\CR$-multiplet -- the supercurrent multiplet of three-dimensional~$\CN=2$ theories with a~$U(1)_R$ symmetry -- and its coupling to linearized new minimal supergravity. (See~\refs{\DumitrescuIU,\ClossetVP} for a thorough discussion.) This will enable us to derive the Killing spinor equations~\ksei\ and~\kseii.

\subsec{The~$\CR$-Multiplet}

Whenever a three-dimensional~$\CN=2$ theory possesses a~$U(1)_R$ symmetry, the conserved~$R$-current~$j_\mu^{(R)}$ resides in a supercurrent multiplet~$\CR_\mu$, together with the supersymmetry currents~$S_{\mu\alpha}, \t S_{\mu\alpha}$ and the energy-momentum tensor~$T_{\mu\nu}$, as well as various other operators. The superfield~$\CR_\mu$ satisfies
\eqn\rmult{\eqalign{& \t D^\beta \CR_{\alpha\beta} = - 4 i \t D_\alpha \CJ^{(Z)}~, \qquad D^\beta \CR_{\alpha\beta} = 4 i D_\alpha \CJ^{(Z)}~,\cr
& D^2 \CJ^{(Z)} = \t D^2 \CJ^{(Z)} = 0~.}}
Here~$\CR_{\alpha\beta} = -2 \gamma^\mu_{\alpha\beta} \CR_\mu$ is the symmetric bi-spinor corresponding to~$\CR_\mu$.  Note that~$\CJ^{(Z)}$ is a real linear multiplet. In components,
\eqn\rmultcomp{\eqalign{& \CR_\mu = j_\mu^{(R)} - i \theta S_\mu - i \t \theta \t S_\mu - (\theta\gamma^\nu\t \theta) \big(2 T_{\mu\nu} + i \ep_{\mu\nu\rho} \d^\rho J^{(Z)}\big) \cr
& \hskip25pt - i \theta \t \theta \big(2 j_\mu^{(Z)} + i \ep_{\mu\nu\rho} \d^\nu j^{(R)\rho} \big) + \cdots~, \cr
& \CJ^{(Z)} = J^{(Z)} - \half \theta \gamma^\mu S_\mu + \half \t \theta \gamma^\mu \t S_\mu + i \theta \t \theta \, T^\mu_\mu - (\theta \gamma^\mu \t \theta) j_\mu^{(Z)} + \cdots~,}}
where the ellipses denote terms that are determined by the lower components. The superfield~$\CJ^{(Z)}$ contains a conserved current~$j_\mu^{(Z)}$ of dimension three, which gives rise to the central charge~$Z$ in the flat space supersymmetry algebra~\flatspacesusy. The scalar~$J^{(Z)}$ is associated with a conserved string current~$ i\ep_{\mu\nu\rho} \d^\rho J^{(Z)}$.

If the theory possesses Abelian flavor symmetries, the~$\CR$-multiplet is not unique. It can be changed by an improvement transformation,
\eqn\imp{\eqalign{& \CR'_{\alpha\beta} = \CR_{\alpha\beta} - \half \big([D_\alpha, \t D_\beta] + [D_\beta, \t D_\alpha]\big) \CJ~,\cr
& \CJ'^{(Z)} = \CJ^{(Z)} - {i \over 2} \t D D\CJ~,}}
where~$\CJ$ is a real linear multiplet, $D^2 \CJ = \t D^2 \CJ = 0$. In this case the Abelian flavor current residing in~$\CJ$ mixes with the~$R$-current. In a superconformal theory, there exists a preferred~$\CR$-multiplet, for which the real linear multiplet~$\CJ^{(Z)}$ in~\rmult\ vanishes. (More precisely, the operators in~$\CJ^{(Z)}$ become redundant.) The corresponding~$U(1)_R$ symmetry belongs to the~$\CN=2$ superconformal algebra.

\subsec{Linearized New Minimal Supergravity}

At the linearized level, the~$\CR$-multiplet couples to the metric superfield~$\CH_\mu$,
\eqn\rh{{\delta \scr L} =  2 \int d^4 \theta \, \CR_\mu \CH^\mu~.}
This term is manifestly invariant under ordinary flat-space supersymmetry transformations with arbitrary constant spinor parameters~$\zeta, \t \zeta$. The linearized supergravity gauge transformations are embedded in superfields~$L_\alpha, \t L_\alpha$ and the metric superfield transforms as follows:
\eqn\htrans{\delta \CH_{\alpha\beta} = \half \big(D_\alpha \t L_\beta - \t D_\beta L_\alpha\big) + \left(\alpha \leftrightarrow \beta\right)~.}
In order for~\rh\ to be gauge invariant, we must impose the constraint
\eqn\lcons{D^\alpha \t D^2 L_\alpha + \t D^\alpha D^2 \t L_\alpha = 0~.}
We can use~\htrans\ to partially fix a Wess-Zumino gauge, in which the metric superfield is given by
\eqn\wzh{\CH_\mu = \half (\theta \gamma^\nu \t \theta) \left(h_{\mu\nu} + B_{\mu\nu}\right) - {i \over 2} \theta \t \theta \, C_\mu - {i \over 2} \theta^2 \t \theta \t \psi_\mu + {i \over 2} {\t \theta}^{\, 2} \theta \psi_\mu + \half \theta^2 \t \theta^{\, 2} \left(A_\mu - V_\mu\right)~,}
where~$V^\mu = -i \ep^{\mu\nu\rho} \d_\nu C_\rho$. Note that~$h_{\mu\nu}, B_{\mu\nu}, C_\mu, A_\mu$ have vanishing~$R$-charge, while the~$R$-charges of~$\psi_\mu$ and~$\t \psi_\mu$ are~$+1$ and~$-1$, respectively.  In conformal supergravity, the~$\CR$-multiplet reduces to a smaller supercurrent with~$\CJ^{(Z)} = 0$. Consequently, the metric superfield~$\CH_\mu$ enjoys more gauge freedom (the constraint~\lcons\ is absent), which allows us to set~$A_\mu - \half V_\mu$ and~$H$ to zero. The combination~$A_\mu - {3 \over2} V_\mu$ remains.

The residual gauge transformations that preserve the form of~$\CH_\mu$ in~\wzh\ take the form
\eqn\resgt{\eqalign{& \delta h_{\mu\nu} = \d_\mu \Lambda^{(h)}_\nu + \d_\nu \Lambda^{(h)}_\mu~, \qquad \delta B_{\mu\nu} = \d_\mu \Lambda^{(B)}_\nu - \d_\nu \Lambda^{(B)}_\mu~,\cr
& \delta C_\mu = \d_\mu \Lambda^{( C )}~, \qquad  \delta A_\mu = \d_\mu \Lambda^{(A)}~,\cr
& \delta \psi_{\mu\alpha} = \d_\mu \ep_\alpha~, \qquad \delta \t \psi_{\mu\alpha} = \d_\mu \t \ep_\alpha~.}}
This identifies~$h_{\mu\nu}$ as the linearized metric, which we normalize so that~$g_{\mu\nu} = \delta_{\mu\nu} + 2 h_{\mu\nu}$, $B_{\mu\nu}$ as a two-form gauge field, $C_\mu$ and~$A_\mu$ as Abelian gauge fields, and~$\psi_{\mu\alpha}, \t \psi_{\mu\alpha}$ as the gravitini. We will often use the gauge-invariant dual field strengths
\eqn\fsdef{\eqalign{& V^\mu = -i \ep^{\mu\nu\rho} \d_\nu C_\rho~, \qquad \d_\mu V^\mu = 0~,\cr
& H = {i \over 2} \ep^{\mu\nu\rho} \d_\mu B_{\nu\rho}~.}}
Since we are working in Euclidean signature, the superfields~$L_\alpha, \t L_\alpha$, and hence the gauge parameters in~\resgt, are complex. In order to retain a sensible geometric description for the supergravity fields, we will demand that the metric~$h_{\mu\nu}$ and the gauge parameter~$\Lambda^{(h)}_\mu$ be real. Although we will generally allow the other supergravity fields to be complex, we will not allow complexified gauge transformations for~$C_\mu$ and~$A_\mu$. Therefore, we also take the gauge parameters~$\Lambda^{( C)}$ and~$\Lambda^{(A)}$ to be real.

In Wess-Zumino gauge, the linearized supergravity-matter couplings~\rh\ are
\eqn\rhcomp{{\delta \scr L} = - T_{\mu\nu} h^{\mu\nu} - \half S_\mu \psi^\mu + \half \t S_\mu \t \psi^\mu+ j_\mu^{(R)} \big(A^\mu - {3 \over 2} V^\mu \big) + j_\mu^{(Z)} C^\mu + J^{(Z)} H~.}
As usual, the metric~$h_{\mu\nu}$ couples to the energy-momentum tensor~$T_{\mu\nu}$, and the gravitini~$\psi_\mu, \t \psi_\mu$ couple to the supersymmetry currents~$S_\mu, \t S_\mu$. Since~$A_\mu$ couples to the~$R$-current~$j_\mu^{( R )}$, we see that~$\Lambda^{(A)}$-transformations in~\resgt\ correspond to local~$R$-transformations. Similarly, the graviphoton~$C_\mu$ gauges the central charge current~$j_\mu^{(Z)}$ and~$\Lambda^{( C )}$-transformations in~\resgt\ correspond to local~$Z$-transformations. The fact that~$H$ couples to~$J^{(Z)}$ means that the two-form~$B_{\mu\nu}$ gauges the string current~$i \ep_{\mu\nu\rho} \d^\rho J^{(Z)}$.

Once we fix Wess-Zumino gauge, the form of~$\CH_\mu$ in~\wzh\ is no longer invariant under conventional flat-space supersymmetry transformations. As usual, this problem can be circumvented by combining these supersymmetry transformations with a gauge transformation~\htrans\ that restores Wess-Zumino gauge. The theory is then invariant under these new supersymmetry transformations, as well as the residual gauge transformations~\resgt. The supersymmetry transformations of the gravitini are thus given by
\eqn\wzgravvar{\eqalign{& \delta \psi_\mu = - i \ep^{\nu\rho\lambda} \d_\nu h_{\rho\mu} \gamma_\lambda \zeta  - 2 i \left(A_\mu - V_\mu\right) \zeta + H\gamma_\mu \zeta + \ep_{\mu\nu\rho} V^\nu \gamma^\rho \zeta + \d_\mu (\cdots)~,\cr
& \delta \t \psi_\mu =- i \ep^{\nu\rho\lambda} \d_\nu h_{\rho\mu} \gamma_\lambda \t \zeta  + 2 i \left(A_\mu - V_\mu\right) \t \zeta + H\gamma_\mu \t \zeta -\ep_{\mu\nu\rho} V^\nu \gamma^\rho \t \zeta + \d_\mu (\cdots)~.}}
Here, the ellipses represent terms that can be absorbed using the residual gauge transformations~\resgt. So far, the spinors~$\zeta, \t \zeta$ in~\wzgravvar\ are constant and parameterize ordinary flat-space supersymmetry transformations, while~$\ep, \t \ep$ in~\resgt\ are arbitrary functions that reflect the residual gauge freedom of the gravitini. As we will see below,  these parameters must be identified in the non-linear theory.

\subsec{Variation of the Gravitino}

In principle, the linearized analysis of the previous subsection constitutes the starting point for a systematic development of the full non-linear supergravity theory. Here we will limit ourselves to explaining why the variation of the gravitini in non-linear supergravity is given by~\gravivar.

In order to pass to non-linear supergravity, we make the supersymmetry transformation parameters~$\zeta, \t \zeta$ spacetime dependent. Under such a transformation, the Lagrangian~${\scr L}_0$ of the original flat-space matter theory is no longer invariant. As usual, its variation is given by
\eqn\deltalzero{\delta {\scr L}_0 = S^\mu \d_\mu \zeta - \t S^\mu \d_\mu \t \zeta~.}
Comparing with~\rhcomp, we see that the linearized supergravity-matter couplings enable us to absorb this variation by a gravitino gauge transformation~\resgt\ with parameters~$\ep = 2 \zeta$ and~$\t \ep  = 2 \t \zeta$. Therefore, the fact that~$\zeta, \t \zeta$ are spacetime dependent requires us to combine supersymmetry transformations with the residual gauge freedom of the gravitini, so that
\eqn\wzgravvarii{\eqalign{& \delta \psi_\mu = 2 \big(\d_\mu \zeta - {i\over 2} \ep^{\nu\rho\lambda} \d_\nu h_{\rho\mu} \gamma_\lambda \zeta\big)  - 2 i A_\mu\zeta + 2i V_\mu \zeta + H\gamma_\mu \zeta + \ep_{\mu\nu\rho} V^\nu \gamma^\rho \zeta~,\cr
& \delta \t \psi_\mu = 2 \big( \d_\mu \t \zeta- {i\over 2} \ep^{\nu\rho\lambda} \d_\nu h_{\rho\mu} \gamma_\lambda \t \zeta \big) + 2 i A_\mu \t \zeta - 2i  V_\mu  \t \zeta + H\gamma_\mu \t \zeta -\ep_{\mu\nu\rho} V^\nu \gamma^\rho \t \zeta~.}}
Note that the terms in parentheses are the linearized covariant derivatives~$\grad_\mu \zeta$ and~$\grad_\mu \t \zeta$.

The transformation laws~\wzgravvarii\ are valid at leading order in an expansion around flat space. The terms that arise at higher orders are of two kinds:
\medskip
\item{1.)} Terms that are needed to render~\wzgravvarii\ fully covariant.
\item{2.)} Additional covariant terms not present in~\wzgravvarii.
\medskip
\noindent
The terms of the first kind have the effect or replacing the terms in parentheses with~$\grad_\mu \zeta$ and~$\grad_\mu \t \zeta$, as well as covariantizing the definition of~$V^\mu$ in~\fsdef, so that~$\grad_\mu V^\mu = 0$. In order to constrain the terms of the second kind, we use dimensional analysis, as well as the covariance of the full non-linear theory under diffeomorphisms and local~$R$-transformations.

The relevant scaling dimensions are given by
\eqn\scaldim{[\psi_\mu] = [\t \psi_\mu] = \half~, \quad [\zeta] = [\t \zeta] = - \half~, \quad [g_{\mu\nu}] = 0~, \quad [A_\mu] = [V_\mu] = [H] = 1~.}
Since all terms in~$\delta \psi_\mu, \delta \t \psi_\mu$ must be proportional to~$\zeta, \t \zeta$, they can contain at most one spacetime derivative, one power of~$A_\mu, V_\mu, H$, or a gravitino bilinear. Therefore, the only possible terms of the second kind are proportional to the gravitini, and hence the transformation laws in the full non-linear theory must take the form in~\gravivar,
\eqn\gravivarbis{\eqalign{& \delta \psi_\mu = 2 \left(\grad_\mu - i A_\mu\right)\zeta + H \gamma_\mu \zeta + 2 i V_\mu \zeta + \ep_{\mu\nu\rho} V^\nu \gamma^\rho \zeta + \cdots~,\cr
& \delta \t \psi_\mu = 2 \left(\grad_\mu + i A_\mu\right) \t \zeta + H \gamma_\mu \t \zeta - 2 i V_\mu \t \zeta - \ep_{\mu\nu\rho} V^\nu \gamma^\rho \t \zeta + \cdots~,}}
where the ellipses denote terms proportional to~$\psi_\mu$ or~$\t \psi_\mu$. Such terms are absent in the rigid limit. Thus, we have determined all relevant terms in the gravitino variations from linearized supergravity. It is also possible to derive these variations by dimensionally reducing the known gravitino variations in four-dimensional new minimal supergravity. This is explained in appendix~D.

\newsec{General Properties of the Killing Spinor Equations}

As explained in the introduction, a given set of supergravity background fields preserves rigid supersymmetry if and only if it is possible to find solutions~$\zeta$ or~$\t \zeta$ of the Killing spinor equations~\ksei\ or~\kseii,
\eqn\ksebis{\eqalign{
& \left(\grad_\mu - i A_\mu\right) \zeta =  -\half H \gamma_\mu\zeta - i V_\mu\zeta -\half \ep_{\mu\nu\rho}V^\nu \gamma^\rho\zeta~,\cr
& \left(\grad_\mu + iA_\mu\right) \t\zeta =  -\half H \gamma_\mu\t\zeta + i V_\mu\t\zeta +\half\ep_{\mu\nu\rho}V^\nu \gamma^\rho\t\zeta~.}}
Note that the second equation can be obtained from the first one by substituting
\eqn\eqsymm{\zeta \rightarrow \t \zeta~, \qquad A_\mu \rightarrow -A_\mu~, \qquad V_\mu \rightarrow -V_\mu~, \qquad H \rightarrow H~.}

In this section we begin our analysis of the equations~\ksebis. We will study them on a smooth, oriented, connected three-manifold~$\CM$, which is endowed with a Riemannian metric~$g_{\mu\nu}$. The background fields~$A_\mu, V_\mu, H$ are generally complex, and the dual graviphoton field strength~$V^\mu$ is covariantly conserved, $\grad_\mu V^\mu = 0$. Since the Killing spinor equations only depend on the graviphoton through~$V^\mu$, they only determine it up to a flat connection.

The spinors~$\zeta$ and~$\t \zeta$ transform as doublets under local~$SU(2)$ frame rotations, and they carry~$R$-charge~$+1$ and~$-1$, respectively. We will only consider conventional~$R$-transformations, so that the real part~$\Re A_\mu$ transforms as a~$U(1)_R$ connection, while the imaginary part~$\Im A_\mu$ is a well-defined one-form. If we denote by~$L$ the line bundle of local~$U(1)_R$ transformations and by~$S$ the spin bundle, then~$\zeta$ and~$\t\zeta$ are sections of~$L \otimes S$ and~$L^{-1} \otimes S$, respectively. If we fix a spin structure on~$\CM$, the bundles~$L$ and~$S$ are well defined, but our discussion only requires a spin$^c$ structure, which exists on every orientable three-manifold. Then only the product bundles~$L \otimes S$ and~$L^{-1} \otimes S$ are well defined.

The equations~\ksebis\ are linear, homogenous, and first-order, with smooth coefficients. Therefore, the solutions have the structure of a complex vector space, which decomposes into solutions~$\zeta$ of~$R$-charge~$+1$ and solutions~$\t \zeta$ of~$R$-charge~$-1$. Importantly, any non-trivial solution is nowhere vanishing on~$\CM$.\foot{To see this, assume that~$\zeta(x_0) = 0$ for some point~$x_0 \in \CM$. We can connect any other point~$x_1 \in \CM$ to~$x_0$ by a smooth curve~$x(s)$, along which~\ksebis\ reduces to an ordinary differential equation of the form~${d \over ds} \zeta (x(s)) = M(s) \zeta (x(s))$, for some smooth matrix-valued function~$M(s)$. The fact that~$\zeta(x_0) = 0$ implies that~$\zeta$ vanishes everywhere on the curve~$x(s)$, and hence at~$x_1$.} This implies that every solution is determined by its value at a single point, and hence there are at most two independent solutions of~$R$-charge~$+1$ and two independent solutions of~$R$-charge~$-1$.

In order to analyze the conditions under which the equations~\ksebis\ admit one or several solutions, it is convenient to use these solutions to construct spinor bilinears. Here we will study such bilinears at a point, making use of Fierz identities but not of the equations~\ksebis.

Given a non-zero spinor~$\zeta \in L \otimes S$, we can define its norm~$|\zeta|^2$, which is positive, and a real, non-vanishing one-form
\eqn\etadef{\eta_\mu={1\over |\zeta|^2}\zeta^\dagger \gamma_\mu \zeta~,}
which satisfies
\eqn\etapr{\eta^\mu\eta_\mu=1~.}
We can use~$\eta_\mu$ to define a tensor
\eqn\amcsform{{\Phi^\mu}_\nu = {\ep^\mu}_{\nu\rho} \eta^\rho~,}
which leads to the following identity:
\eqn\acs{{\Phi^\mu}_\rho{\Phi^\rho}_\nu=-{\delta^\mu}_\nu + \eta^\mu \eta_\nu~.}
These formulas imply that~$\eta_\mu$ defines an almost contact metric structure (ACMS) on~$\CM$. Here we will only summarize its essential properties. A more detailed discussion can be found in appendix~B.

An ACMS is the odd-dimensional analogue of an almost Hermitian structure. In three dimensions, it is equivalent to a reduction of the tangent bundle structure group to~$U(1)$. It follows from~\amcsform\ that~${\Phi^\mu}_\nu$ has rank two and that~${\Phi^\mu}_\nu \eta^\nu = 0$. A vector~$X^\mu$ or a one-form~$\Omega_\mu$ is called holomorphic if~${\Phi^\mu}_\nu X^\nu = i X^\mu$ or~$\Omega_\mu {\Phi^\mu}_\nu = i \Omega_\nu$. (This implies that both are orthogonal to~$\eta_\mu$.) Their complex conjugates~$\b X^\mu$ and~$\b \Omega_\mu$ are called anti-holomorphic. Since~$\eta_\mu$ is defined in terms of~$\zeta$ via~\etadef, it follows that~$X^\mu$ is holomorphic if and only if~$X^\mu \gamma_\mu \zeta=0$,\foot{If~$X^\mu$ is holomorphic, then~$\zeta^\dagger \gamma_\mu \gamma_\nu \zeta X^\nu=0$. Multiplying by~$\b X^\mu$, we find that~$|X^\nu \gamma_\nu \zeta|^2=0$, so that $X^\nu\gamma_\nu \zeta=0$. Conversely, if~$X^\nu \gamma_\nu \zeta=0$ we can multiply by~$\zeta^\dagger \gamma^\mu$ to obtain~${\Phi^\mu}_\nu X^\nu=i X^\mu$.} and that~$\eta_\mu \gamma^\mu \zeta=\zeta$.

We can also use~$\zeta$ to construct a nowhere vanishing one-form of~$R$-charge~$2$,
\eqn\defpbp{P_\mu=\zeta\gamma_\mu \zeta~,}
which satisfies
\eqn\pahol{P_\mu {\Phi^\mu}_\nu=-i P_\nu~.}
Therefore~$P_\mu$ is an anti-holomorphic one-form.

We can repeat the preceding discussion for a nowhere vanishing spinor~$\t \zeta \in L^{-1} \otimes S$. This gives rise to a nowhere vanishing one-form~$\t \eta_\mu = {1 \over |{\t \zeta}|^2} \t \zeta^\dagger \gamma_\mu \t \zeta $, which defines an ACMS and a nowhere vanishing one-form~$\t P_\mu = \t \zeta \gamma_\mu \t \zeta$ of~$R$-charge~$-2$ that is anti-holomorphic with respect to this ACMS.

\newsec{Manifolds Admitting One Supercharge}

In this section we establish necessary and sufficient conditions for the existence of a solution~$\zeta$ of the Killing spinor equation~\ksei,
\eqn\kseibis{\left(\grad_\mu - i A_\mu\right) \zeta =  -\half H \gamma_\mu\zeta - i V_\mu\zeta -\half \ep_{\mu\nu\rho}V^\nu \gamma^\rho\zeta~.}
The presence of a solution imposes a certain integrability condition on the ACMS defined in the previous section. Conversely, a solution exists whenever~$\CM$ admits such an integrable ACMS. The corresponding results for a solution~$\t \zeta$ of~\kseii\ are easily obtained using~\eqsymm.

\subsec{Restrictions Imposed by~$\zeta$}

In~\etadef\ and~\defpbp\ we used~$\zeta$ to construct two nowhere vanishing bilinears~$\eta_\mu$ and~$P_\mu$, and we saw that~$\eta_\mu$ and~${\Phi^\mu}_\nu = {\ep^\mu}_{\nu\rho} \eta^\rho$ define an ACMS on~$\CM$. Here we will use the fact that~$\zeta$ satisfies the Killing spinor equation~\kseibis\ to study the derivatives of these bilinears. This leads to an integrability condition for the ACMS defined by~$\eta_\mu$. It also allows us to constrain the background fields~$A_\mu, V_\mu, H$, although it does not determine them completely. This is because the equation~\kseibis\ remains invariant under the following shifts:
\eqn\shiftbck{\eqalign{&V^\mu \rightarrow V^\mu +U^\mu+\kappa \eta^\mu,\cr
& H\rightarrow H+i \kappa ~,\cr
&A_\mu\rightarrow A_\mu+ {3\over 2}(U_\mu+ \kappa \eta_\mu)~,}}
where the complex scalar~$\kappa$ and the vector~$U^\mu$ must satisfy
\eqn\lucond{{\Phi^\mu}_\nu U^\nu=i U^\mu~, \qquad \grad_\mu (U^\mu+ \kappa \eta^\mu)=0~.}
The solution~$\zeta$ completely determines the background fields up to these shifts.

First, we use~\kseibis\ to compute the derivative of~$\eta_\mu$,
\eqn\gradeta{\eqalign{\grad_\mu \eta_\nu =~& \half\left(H + \b H\right) \left(\eta_\mu \eta_\nu - g_{\mu\nu}\right) + {i \over 2} \left(H - \b H\right) \Phi_{\mu\nu}+ {i \over2} g_{\mu\nu} \eta_\rho \big(V^\rho - \b V^\rho\big) \cr
& - {i \over 2} \eta_\mu \big(V_\nu - \b V_\nu\big)+ \half \Phi_{\mu\nu} \eta_\rho \big(V^\rho + \b V^\rho\big) + \half \eta_\mu \Phi_{\nu\rho} \big(V^\rho + \b V^\rho\big)~.}}
This fixes~$V^\mu$ and $H$, up to the shifts in~\shiftbck,
\eqn\auxsol{V^\mu= {\ep}^{\mu \nu\rho}\d_\nu \eta_\rho~,\qquad H=-{1\over 2} \grad_\mu \eta^\mu +{i\over 2} \ep^{\mu\nu\rho} \eta_\mu \d_\nu \eta_\rho~.}
Substituting back into~\gradeta, we find that~$\eta_\nu$ must satisfy the integrability condition
\eqn\etacond{\grad_\mu \eta_\nu + \grad_\nu \eta_\mu = \left(g_{\mu\nu} - \eta_\mu \eta_\nu\right) \grad_\rho \eta^\rho + \eta_\mu \eta^\rho \grad_\rho \eta_\nu + \eta_\nu \eta^\rho \grad_\rho \eta_\mu~.}
This condition can be succinctly written in terms of~${\Phi^\mu}_\nu$,
\eqn\liecondd{{\Phi^\mu}_\nu \left(\CL_\eta {\Phi^\nu}_\rho\right)=0~.}
Here~$\CL_\eta$ is the Lie derivative along~$\eta^\mu$ (see appendix~A). A similar analysis in~\KlareGN\ found the integrability condition~$\ep^{\mu\nu\rho} P_\mu \d_\nu P_\rho = 0$, which is locally equivalent to~\etacond.

The integrability condition~\liecondd\ is analyzed in appendix~B. It is equivalent to a covering of~$\CM$ by adapted coordinate charts~$(\tau, z, \b z)$, with real~$\tau$ and complex~$z$, which satisfy the following properties:

\medskip

\item{1.)} Two overlapping adapted charts~$(\tau, z, \b z)$ and~$(\tau', z', \b z')$ are related by
\eqn\holtrans{\tau' = \tau + t(z, \b z)~, \qquad z' = f(z)~,}
where~$t(z,\b z)$ is real and~$f(z)$ is holomorphic.

\item{2.)} In an adapted chart~$(\tau, z , \b z)$, the vector $\eta^\mu$ is given by
\eqn\etachart{\eta^\mu \d_\mu = \d_\tau~.}

\item{3.)} In an adapted chart~$(\tau, z , \b z)$, a holomorphic one-form~$\Omega_\mu$ is given by
\eqn\holom{\Omega_\mu dx^\mu = \omega(\tau, z,\b z) dz~.}
Changing adapted coordinates as in~\holtrans\ leads to the transformation law
\eqn\ochange{\omega' (\tau', z', \b z') = {1 \over f'(z)} \, \omega (\tau, z, \b z)~.}
The line bundle~$\CK$ of holomorphic one-forms thus has holomorphic transition functions.

\item{4.)} In an adapted chart~$(\tau, z , \b z)$, the metric takes the form
\eqn\acmsmets{ds^2=\big(d\tau+ h(\tau,z,\b z) dz+\b h(\tau,z,\b z) d \b z\big)^2+c(\tau,z,\b z)^2 dz d\b z~,}
where~$h(\tau, z, \b z)$ and~$c(\tau, z, \b z)$ are complex and real, respectively.

\medskip

\noindent These properties suggest that an ACMS that satisfies the integrability condition~\liecondd\ constitutes a natural three-dimensional analogue of an integrable complex structure.  Indeed, such an ACMS is equivalent to a structure known as a transversely holomorphic foliation, which has been studied in the mathematical literature (see~\ClossetVRA\ for additional details and references).

It is convenient to define a connection~$\hat \grad_\mu$ that satisfies~$\hat \grad_\mu g_{\nu\rho} = 0$ and~$\hat \grad_\mu \eta_\nu = 0$.\foot{Similarly, there are metric-compatible connections in even dimensions that preserve a given complex structure. Such connections played an important role in~\refs{\DumitrescuHA,\DumitrescuAT}.} This can be accomplished by replacing the usual spin connection~$\omega_{\mu\nu\rho}$ by
\eqn\hatom{\hat \omega_{\mu\nu\rho} = \omega_{\mu\nu\rho} + \eta_\rho \grad_\mu \eta_\nu - \eta_\nu \grad_\mu \eta_\rho + 2 W_\mu \Phi_{\nu\rho}~, \qquad W_\mu=-{1\over 4} \eta_\mu \ep^{\nu\rho\lambda}\eta_\nu \d_\rho \eta_\lambda~.}
Unless~$\eta_\mu$ is covariantly constant, the connection~$\hat \grad_\mu$ has torsion (see appendix~B). Since this connection preserves~$\eta_\mu$, and hence the associated ACMS, its holonomy is contained in~$U(1)$. In terms of~$\hat \grad_\mu$, we can rewrite the Killing spinor equation~\kseibis\ as follows:
\eqn\difcob{\big(\hat \grad_\mu-i \hat A_\mu\big)\zeta=0~,\qquad   \hat A_\mu=  A_\mu-\half \big(2 {\delta_\mu}^\nu - i {\Phi_\mu}^\nu\big) V_\nu +{i\over 2} \eta_\mu H- W_\mu~. }
Note that~$\hat A_\mu$ differs from~$A_\mu$ by a well-defined one-form and that it is not affected by the shifts in~\shiftbck. Since the holonomy of~$\hat \grad_\mu$ is contained in~$U(1)$, we can twist it away by adjusting~$\hat A_\mu$, so that~$\zeta$ transforms as a scalar on~$\CM$. Conversely, the twisting procedure allows us to solve for~$\zeta$ on any manifold that admits an ACMS satisfying~\liecondd. We will return to this point below. (See~\DumitrescuHA\ for a related discussion in four dimensions.)

In order to determine~$\hat A_\mu$, and hence~$A_\mu$, we consider $P_\mu=\zeta\gamma_\mu \zeta$, which has~$R$-charge~$2$. Note that~$P_\mu$ locally determines $\zeta$ up to a sign. It follows from~\pahol\ that~$P_\mu$ is a nowhere vanishing section of~$L^2 \otimes \b \CK$, where~$\b \CK$ is the line bundle of anti-holomorphic one-forms. Therefore, the bundle~$L^2 \otimes \b \CK$ is trivial, and we can identify~$L = \big(\b \CK \big)^{- \half}$, up to a trivial line bundle. To make this explicit, we work in an adapted chart~$(\tau, z, \b z)$ and define~$p = P_{\b z}$ as in~\holom. If we define
\eqn\defsc{s={1\over \sqrt{2}} p g^{-{1\over 4}}~,}
then~$s$ transforms by a phase under under a change~\holtrans\ of adapted coordinates,
\eqn\chch{s'\big(\tau', z',\b z'\big)=\left({f'(z)\over\b f'(\b z)} \right)^{\ha} s\big(\tau,z , \b z\big ) ~.}
Locally, these phase rotations can be compensated by appropriate~$R$-transformations. Under these combined transformations~$s$ is a scalar. As we will see in the next subsection, this scalar determines the solution~$\zeta$.

We will now solve for~$\hat A_\mu$ in terms of~$s$. It follows from~\difcob\ that
\eqn\devp{\big(\hat \grad_\mu-2 i \hat A_\mu\big) p=0~,}
so that~$\hat A_\mu = - {i \over 2} \hat \grad_\mu \log p$. In an adapted chart, the metric is given by~\acmsmets, and thus
\eqn\derp{\eqalign{&\hat \grad_\tau s= \d_\tau s~,\cr
& \hat \grad_z s=\d_z s +{s\over 4}\big( \d_z -\eta_z \d_\tau\big) \log g~,\cr
&\hat \grad_{\b z} s=\d_{\b z} s-{s\over 4}\big(\d_{\b z}- \eta_{\b z} \d_\tau\big) \log g~.}}
Using~\defsc, we therefore find that~$\hat A_\mu$ is given by
\eqn\Asol{\hat A_\mu = {1\over 8} {\Phi_\mu}^\nu \d_\nu \log g  -{i\over 2} \d_\mu \log s~.}
Note that this expression is only valid in an adapted chart and does not transform covariantly under arbitrary coordinate changes.\foot{If we compute the field strength~$\d_\mu \hat A_\nu - \d_\nu \hat A_\mu$ from~\Asol, we obtain a fully covariant expression. This determines~$\hat A_\mu$ up to a flat connection.} However, under a change~\holtrans\ of adapted coordinates, the real part of~$\hat A_\mu$ transforms like a~$U(1)$ gauge field, while its imaginary part is invariant.

\subsec{Solving for~$\zeta$}

We will now solve the Killing spinor equation~\kseibis\ on every thee-manifold~$\CM$ that admits an ACMS satisfying the integrability condition~\liecondd. Up to the shifts~\shiftbck, the background fields~$V_\mu, H$ are given by~\auxsol, while~$A_\mu$ is given by~\difcob\ and~\Asol.

In order to write down the solution, we work in adapted coordinates~$(\tau, z, \b z)$, so that the metric is given by~\acmsmets, and we use an orthonormal frame~$e^1, e^2, e^2$ defined by\foot{This definition is such that the usual orientation~$e^1 \wedge e^2 \wedge e^3$ corresponds to~${\Phi^z}_z = - {\Phi^{\b z}}_{\b z} = i$ (see appendix B).}
\eqn\frmad{e^1= \eta~, \qquad e^2 - i e^3 = c\left(\tau, z, \b z\right) dz~,\qquad e^2 + i e^3 = c\left(\tau, z, \b z\right) d\b z~.}
In this frame, the solution~$\zeta$ takes the form
\eqn\zetasol{\zeta_\alpha =\sqrt {s\left(\tau, z , \b z\right)} \pmatrix{1 \cr 0}~,}
where~$s$ is the same nowhere vanishing function as in~\defsc. Since~$s$ transforms as a scalar under changes of adapted coordinates followed by an appropriate~$R$-transformation, we see that~$\zeta$ also transforms as a scalar. Therefore, it gives rise to a globally well-defined scalar supercharge on~$\CM$.

\newsec{Manifolds Admitting Multiple Supercharges}

 In this section we will consider manifolds that admit more than one solution of the Killing spinor equations~\ksei\ and~\kseii, focusing on two supercharges of opposite~$R$-charge and four supercharges. (We do not discuss the case of two solutions with equal~$R$-charge.) The first case was already discussed in~\DumitrescuHA. Here we emphasize some important features that allow us to make contact with the various known squashings of the three-sphere.

\subsec{Two Supercharges of Opposite~$R$-Charge}

Given solutions~$\zeta$ of~\ksei\ and~$\t \zeta$ of~\kseii, we can define a nowhere vanishing vector
\eqn\defkv{K^\mu= \zeta \gamma^\mu \t \zeta~.}
Using the Killing spinor equations, we find that it is a Killing vector,
\eqn\killcheck{\grad_\mu K_\nu+\grad_\nu K_\mu=0~.}
Since~$K_\mu$ is complex, it may give rise to two independent isometries. This case is very restrictive, and we do not discuss it here. Instead, we assume that~$K^\mu$ generates a single isometry. In this case we can normalize the spinors, so that
\eqn\solprop{\t \zeta={\Omega \over|\zeta|^2} \, \zeta^{\dagger}~,}
for some positive scalar function~$\Omega$. With this normalization~$K^\mu$ is real. It follows from~\solprop\ that~$\Omega^2 = K^\mu K_\mu$, so that~$\Omega$ is invariant under the Killing vector~$K^\mu$. The ACMS's corresponding to~$\zeta$ and~$\t \zeta$ are now simply related to~$K_\mu$,
\eqn\relacms{\eta_\mu=-\t \eta_\mu =  \Omega^{-1} K_\mu~.}
Since~$K^\mu$ is Killing, it follows that the ACMS's satisfy the integrability condition~\liecondd.

Since~$K^\mu$ is a real, nowhere vanishing Killing vector, it is natural to introduce coordinates~$(\psi, z, \b z)$, such that~$K = \d_\psi$. In these coordinates, the metric takes the form
\eqn\metriso{ds^2=\Omega^2(z, \b z) \big(d\psi+a\left(z,\bar z\right)dz+ \b a\left(z, \b z\right) d \b z\big)^2+ c(z ,\b z)^2 dz d\b z~.}
It follows from~\relacms\ that~$\eta_\mu dx^\mu = \Omega(z, \b z) \big(d\psi+a\left(z,\bar z\right)dz+ \b a\left(z, \b z\right) d \b z\big)$. The coordinates~$(\tau, z, \b z)$ adapted to~$\eta_\mu$ are simply related to~$(\psi, z, \b z)$ via~$d \tau = \Omega d\psi$.

In the previous section, we have seen that a solution~$\zeta$ fixes the background fields~$A_\mu, V_\mu, H$ according to~\auxsol, \difcob, and~\Asol, up to shifts~\shiftbck\ by~$\kappa$ and~$U^\mu$. Using~\eqsymm, we can similarly obtain the background fields corresponding to~$\t \zeta$, up to shifts by~$\t \kappa$ and~$\t U^\mu$. If both solutions are present simultaneously, we must ensure that the corresponding background fields are consistent. Using~\relacms, we find that the background fields are given by the same formulas as before, but the freedom of performing shifts~\shiftbck\ is constrained: \eqn\shiftcons{U^\mu = \t U^\mu = 0~, \qquad \kappa = - \t \kappa~, \qquad K^\mu \d_\mu \kappa = 0~.}
The fact that~$\kappa$ is invariant along~$K^\mu$ ensures that~$V_\mu, H$ are invariant under~$K^\mu$, while~$A_\mu$ is only invariant up to a (complexified) gauge transformation. (See~\DumitrescuHA\ for a related discussion in four dimensions.)

Having determined the form of the background fields, we can write down the explicit solution for~$\zeta$ and~$\t \zeta$ in the adapted frame~\frmad,
\eqn\zetasol{\zeta_\alpha =\sqrt{s(\psi, z , \b z)} \pmatrix{1 \cr 0}~,\qquad \t \zeta_\alpha ={\Omega(z, \b z) \over   \sqrt{s(\psi, z , \b z)}} \pmatrix{0 \cr 1}~.}
As in the previous subsection, $\zeta$ and~$\t \zeta$ are globally well-defined solutions on all of~$\CM$, and hence two supercharges of opposite~$R$-charge exist on any manifold admitting a real, nowhere vanishing Killing vector. This covers all previously known examples possessing two such supercharges with real~$K^\mu$. However, there are cases (such as the~$S^2 \times S^1$ background of~\ImamuraSU, which preserves four supercharges and is discussed below) in which~$K^\mu$ is genuinely complex and leads to two independent real Killing vectors. We have not analyzed such backgrounds in detail.

It is important to emphasize that the real Killing vector~$K^\mu$ discussed above need not descend from a single~$U(1)$ isometry of~$\CM$. When~$\CM$ is compact, we distinguish the following cases:
\medskip
\item{1.)} If the orbits of~$K^\mu$ are closed, they give rise to a circle bundle over a Riemann surface, i.e. a Seifert manifold. For instance, the $SU(2)\times U(1)$ invariant squashed sphere of~\HamaEA\ is of this kind.

\item{2.)} If the orbits of~$K^\mu$ are not closed, the manifold necessarily admits additional isometries. It can be shown that~$K^\mu$ is a linear combination, with incommensurate coefficients, of two commuting Killing vectors with compact orbits (see for instance~\Isenberg). In this case~$\CM$ is still a Seifert manifold, but~$K^\mu$ no longer points along the fiber. For example, this occurs on the~$U(1) \times U(1)$ invariant squashed spheres of~\refs{\HamaEA, \MartelliFU}.

\subsec{Four Supercharges}

Here we study manifolds admitting two independent solutions of~\ksei\ and two independent solutions of~\kseii. Given any solution~$\zeta$ of~\ksei, we can use the fact that~$\ep_{\mu\nu\rho} [\grad^\nu, \grad^\rho] \zeta={i\over 2}(2 R_{\mu\nu}-R g_{\mu\nu})\gamma^\nu\zeta$ to obtain the following integrability condition:
\eqn\intcondi{\eqalign{& {i\over 2}(2 R_{\mu\nu}-R g_{\mu\nu})\gamma^\nu \zeta - 2i\ep_{\mu\nu\rho}\grad^\nu (A^\rho-V^\rho)\zeta \cr
& \hskip34pt - \gamma^\nu\zeta \big(\ep_{\mu\nu\rho}(\grad^\rho+i V^\rho) H-\grad_\nu V_\mu -i H^2 g_{\mu \nu} -i V_\mu V_\nu\big) = 0~.}}
The corresponding integrability condition for a solution~$\t \zeta$ of~\kseii\ follows from~\eqsymm. In the presence of four independent supercharges, these integrability conditions imply
\eqn\intcond{\eqalign{& \d_\mu H = 0~,\cr
& \d_\mu (A_\nu - V_\nu) - \d_\nu (A_\mu - V_\mu) = 0~,\cr
& \grad_\mu V_\nu = - i H \ep_{\mu\nu\rho} V^\rho~,\cr
& R_{\mu\nu} = - V_\mu V_\nu + g_{\mu\nu} (V^\rho V_\rho + 2 H^2)~.}}
Thus, $H$ is constant, $A_\mu=V_\mu$ up to a flat connection, and~$V_\mu$ is a Killing vector of constant norm~$v^2 = V^\mu V_\mu$.

We will now analyze the various cases that arise as a consequence of the integrability conditions~\intcond:
\medskip
\medskip
\item{1.)} If~$V_\mu = 0$, then~$R_{\mu\nu} = 2 H^2 g_{\mu\nu}$. Since the metric is real, $H$ is either real or purely imaginary. Therefore, $\CM$ has constant sectional curvature, and hence it is locally isometric to~$S^3, T^3$, or~$H^3$, depending on whether~$H$ is purely imaginary, zero, or real, respectively. When~$H$ is purely imaginary, we recover the round~$S^3$ of~\refs{\KapustinKZ\JafferisUN-\HamaAV}, see also~\FestucciaWS.

\item{2.)} If~$V_\mu \neq 0$ but~$H = 0$, then~$V_\mu$ is covariantly constant. It follows that~$V_\mu$, and hence~$v$, is either real or purely imaginary. Therefore, $\CM$ is locally isometric to~$\R \times \Sigma$, where~$V_\mu$ points along~$\R$ and~$\Sigma$ is a surface of constant curvature~$R^{(\Sigma)} = 2 v^2$. Hence,~$\Sigma$ is locally isometric to~$S^2, T^2$, or~$H^2$, depending on whether~$v$ is purely imaginary, zero, or real, respectively. As in four dimensions~\FestucciaWS, we can choose~$A_\mu = V_\mu$ and compactify~$\R$ to~$S^1$. When~$\Sigma = S^2$, this background can be used to compute a supersymmetric index~\ImamuraSU.

\item{3.)} If both~$V_\mu \neq 0$ and~$H \neq 0$, then~$H$ is purely imaginary and~$V_\mu$ is either real or purely imaginary. The nowhere vanishing Killing vector~$V_\mu$ then determines a fibration over a surface~$\Sigma$ with a metric~$g_{\mu\nu}^{(\Sigma)}$ of constant curvature,
\eqn\metbs{g^{(\Sigma)}_{\mu\nu}= g_{\mu\nu} -{1\over v^{2}} V_\mu V_\nu~, \qquad R^{(\Sigma)} = 2 \left(v^2 +4 H^2\right)~.}
Therefore, $\Sigma$ is locally isometric to~$S^2, T^2$, or $H^2$:
\medskip
\item{3a.)} If~$\Sigma$ is isometric to a round~$S^2$ of radius ${r \over 2}$, we can introduce coordinates~$(\psi, \theta, \phi)$, where~$0 \leq \theta \leq \pi$ and~$\phi  \sim \phi + 2\pi$ are the usual angular coordinates on~$S^2$. The metric and the background fields are then given by
\eqn\metsqst{\eqalign{&ds^2= {r^2 \over 4} \Big( h^2 \big(d\psi+2  \sin^2 {\theta \over 2} \, d\phi\big)^2+ (d\theta^2+\sin^2 \theta \, d\phi^2)\Big)~,\cr
& V^\mu \d_\mu={2 v\over h r}\d_\psi~, \qquad H= {ih\over  r}~,\qquad v^2={4 \left(h^2-1\right) \over r^2}~, \qquad h \in \R - \{0\}~.}}
Here~$\psi$ is an angular coordinate with periodicity~$\psi \sim \psi + 4 \pi$, which parameterizes a Hopf fibration over~$S^2$. This metric describes a squashed three-sphere~$S^3_b$, which preserves an~$SU(2)\times U(1)$ isometry. (In the limit~$h = \pm 1$, we obtain a round~$S^3$ of radius~$r$.) It is convenient to define a squashing parameter~$b$ via
\eqn\bdef{h={1\over 2}\left(b+{1 \over b}\right)~.}
Supersymmetric theories with four supercharges on~$S^3_b$ were constructed in~\refs{\ImamuraUW,\ImamuraWG}. We will make use of this supersymmetric background in section~8.

\item{3b.)} If~$\Sigma$ is flat, we can introduce coordinates~$(\psi, \rho, \phi)$, such that~$\rho \geq 0$ and~$\phi \sim \phi + 2\pi$ are polar coordinates on~$\Sigma$. The metric and the background fields are then given by
\eqn\metsqf{\eqalign{&ds^2= {r^2 \over 4} \Big(h^2 (d\psi+ \half \rho^2 d\phi\big)^2+ (d\rho^2+\rho^2 d\phi^2)\Big)~,\cr
& V^\mu \d_\mu={4 \over r^2} \d_\psi~, \qquad H= {ih\over r}~,\qquad v^2={4 h^2\over r^2}~, \qquad h \in \R - \{0\}~.}}

\item{3c.)} If~$\Sigma$ is isometric to an~$H^2$ of radius~$r \over 2$, we can use coordinates~$(\psi, \rho, \phi)$ with~$\rho \geq 0$ and~$\phi \sim \phi + 2\pi$ to express the metric and the background fields as follows:
\eqn\metsqht{\eqalign{&ds^2= {r^2 \over 4} \Big(h^2\big(d\psi+2  \sinh^2 {\rho \over 2} \, d\phi\big)^2+ (d\rho^2+\sinh^2 \rho \, d\phi^2)\Big)~,\cr
& V^\mu \d_\mu = {2 v \over h r} \d_\psi~, \qquad H=  {ih\over r}~,\qquad v^2={4(h^2+1)\over r^2}~, \qquad h \in \R -\{ 0 \}~.}}

\newsec{Supersymmetry Multiplets and Lagrangians}

In this section, we discuss the general form of the rigid supersymmetry algebra and its multiplets. This enables us to construct supersymmetric Lagrangians on any three-manifold that admits one or several solutions of the Killing spinor equations~\ksei\ and~\kseii.

\subsec{The Supersymmetry Algebra}

Whenever a given three-manifold admits rigid supersymmetry, the corresponding supersymmetry algebra is obtained by taking the rigid limit of the appropriate algebra of supergravity transformations. Since a suitably gauge-fixed component formulation of new minimal supergravity in three dimensions is not available, we will postulate the corresponding rigid supersymmetry algebra and subject it to various consistency checks.

Given solutions~$\zeta, \eta$ of~$\ksei$ and solutions~$\t \zeta, \t \eta$ of~\kseii, we take the rigid supersymmetry algebra to be
\eqn\rigidsalg{\eqalign{& \{ \delta_\zeta, \delta_{\t \zeta}\} \varphi_{(\fr,\fz)} = - 2i \big( \CL'_{K} \varphi_{(\fr,\fz)}  +  \zeta \t \zeta \left(\fz - \fr H\right) \varphi_{(\fr,\fz)} \big)~, \qquad K^\mu = \zeta \gamma^\mu \t \zeta~,\cr
& \{ \delta_\zeta, \delta_\eta\} \varphi_{(\fr,\fz)} = 0~, \qquad \{ \delta_{\t \zeta}, \delta_{\t \eta}\} \varphi_{(\fr,\fz)}= 0~.}}
Here~$\varphi_{(\fr,\fz)}$ is a field of arbitrary spin, $R$-charge~$\fr$, and central charge~$\fz$. We use~$\CL'_K$ to denote a modified Lie derivative along~$K^\mu$, which is covariant under local~$R$- and~$Z$-transformations,
\eqn\hatliedef{ \CL'_K \varphi_{(\fr,\fz)} = \CL_K \varphi_{(\fr,\fz)} - i \fr K^\mu \Big(A_\mu - \half V_\mu\Big) \varphi_{(\fr,\fz)} - i \fz K^\mu C_\mu \varphi_{(\fr,\fz)}~.}
The algebra~\rigidsalg\ passes several consistency checks:
\medskip
\item{1.)} It is covariant under diffeomorphisms, as well as local~$R$- and~$Z$-transformations.
\item{2.)} It reduces to the usual~$\CN=2$ supersymmetry algebra~\flatspacesusy\ when the metric is flat and the other supergravity background fields vanish.  Note that~$z$ is the eigenvalue of the central charge operator~$Z$ in~\flatspacesusy\ on the field~$\varphi_{(r, z)}$.
\item{3.)} It can be derived from the most general ansatz consistent with~$1.)$ and~$2.)$, as well as dimensional analysis, by demanding that the algebra close whenever the spinor parameters satisfy~\ksei\ and~\kseii.
\item{4.)} It can be obtained from the rigid limit of new minimal supergravity in four dimensions by a twisted dimensional reduction. This is discussed in appendices~C and~D.
\medskip

\subsec{Supersymmetry Multiplets}

In this subsection, we realize the algebra~\rigidsalg\ on a general multiplet whose bottom component is a complex scalar. By imposing constraints, we obtain the transformation rules for chiral, anti-chiral, and real linear multiplets. Using these multiplets, we construct general supersymmetric Lagrangians. Finally, we present the multiplication rules for general multiplets.

Consider a general multiplet~$\CS$, whose bottom component~$C$ is a complex scalar of~$R$-charge~$\fr$ and central charge~$\fz$. Such a multiplet has~$16+16$ independent bosonic and fermionic components,
\eqn\sdef{\CS= (C, \chi_\alpha, \t \chi_\alpha, M, \t M, a_\mu, \sigma, \lambda_\alpha, \t \lambda_\alpha, D)~.}
All components of~$\CS$ carry central charge~$\fz$, while the~$R$-charges relative to the bottom component are given by~$(0, -1, 1, -2, 2, 0, 0, 1, -1, 0)$. The general multiplet~\sdef\ is the curved-space analogue of an unconstrained complex superfield in flat space,
\eqn\sfieldflat{\eqalign{\CS =~& C + i \theta \chi + i \t \theta \t \chi + {i \over 2} \theta^2 M + {i \over 2} \t \theta^{\,2} \t M + (\theta \gamma^\mu \t \theta) a_\mu - i \theta \t \theta \sigma \cr
& + i \t \theta^{\, 2} \theta \Big(\lambda - { i \over 2} \gamma^\mu \d_\mu \t \chi\Big) - i \theta^2 \t \theta \Big(\t \lambda + {i \over 2} \gamma^\mu \d_\mu\chi\Big) - \half \theta^2 \t \theta^{\,2} \Big(D + \half \d^2 C\Big)~.}}

The supersymmetry transformation rules for the components of~$\CS$ are given by
\eqn\scomptrans{\eqalign{& \delta C = i \zeta \chi + i \t \zeta \t \chi~,\cr
& \delta \chi = \zeta M - \t \zeta \left(\sigma + \left(\fz - \fr H\right) C\right) - \gamma^\mu \t \zeta \big(D_\mu C + i a_\mu\big)~,\cr
& \delta \t \chi = \t \zeta \t M - \zeta \left(\sigma - \left(\fz - \fr H\right) C\right) - \gamma^\mu \zeta \big(D_\mu C -i a_\mu \big)~,\cr
& \delta M = - 2 \t \zeta \t \lambda + 2 i \left(\fz - \left(\fr - 2\right) H\right) \t \zeta \chi - 2 i D_\mu \big(\t \zeta \gamma^\mu \chi\big)~,\cr
& \delta \t M = 2 \zeta \lambda - 2 i \left(\fz - \left(\fr + 2\right) H\right) \zeta \t \chi - 2 i D_\mu \big(\zeta \gamma^\mu \t \chi\big)~,\cr
& \delta a_\mu = - i \big(\zeta \gamma_\mu \t \lambda + \t \zeta \gamma_\mu \lambda\big) + D_\mu \big(\zeta \chi - \t \zeta \t \chi\big)~,\cr
& \delta \sigma = - \zeta \t \lambda + \t \zeta \lambda + i \left(\fz - \fr H\right) \big(\zeta \chi - \t \zeta \t \chi\big)~,\cr
& \delta \lambda = i \zeta \left(D + \sigma H\right) - i \ep^{\mu\nu\rho} \gamma_\rho \zeta \, D_\mu a_\nu - \gamma^\mu \zeta \left(\left(\fz - \fr H\right) a_\mu + i D_\mu \sigma - V_\mu \sigma\right)~,\cr
&\delta \t \lambda = - i \t \zeta \left(D + \sigma H\right) - i \ep^{\mu\nu\rho} \gamma_\rho \t \zeta \, D_\mu a_\nu + \gamma^\mu \t \zeta \left(\left(\fz - \fr H\right) a_\mu + i D_\mu \sigma + V_\mu \sigma\right)~,\cr
&\delta D = D_\mu \big( \zeta \gamma^\mu \t \lambda - \t \zeta \gamma^\mu \lambda \big) - i V_\mu \big(\zeta \gamma^\mu \t \lambda + \t \zeta \gamma^\mu \lambda\big) - H \big(\zeta \t \lambda - \t \zeta \lambda\big) \cr
& \hskip25 pt + \left(\fz - \fr H\right) \Big(\zeta \t \lambda + \t \zeta \lambda - i H \big(\zeta \chi - \t \zeta \t \chi\big)\Big) + {i \fr \over 4} \left(R - 2V^\mu V_\mu - 6 H^2\right)\big(\zeta \chi - \t \zeta \t \chi\big)~.}}
Here~$R$ is the Ricci scalar and we have defined a covariant derivative
\eqn\dthree{D_\mu  = \grad_\mu - i \fr \Big(A_\mu - \half V_\mu\Big) - i \fz C_\mu~.}
It can be checked that the transformation rules~\scomptrans\ realize the algebra~\rigidsalg\ whenever the spinors~$\zeta, \eta, \t \zeta, \t \eta$ satisfy the Killing spinor equations~\ksei\ and~\kseii. To show this, it is convenient to use the integrability condition~\intcondi, as well as the Killing spinor equations themselves. Alternatively, the transformation rules~\scomptrans\ can be obtained by appropriately reducing the supersymmetry transformations corresponding to the rigid limit of four-dimensional new minimal supergravity (see appendices~C and~D).

We can obtain other useful supersymmetry multiplets by starting with the general multiplet~\sdef\ and imposing constraints:

\medskip

\item{1.)} {\it Chiral Multiplet:} Imposing~$\t \chi_\alpha = 0$ leads to a chiral multiplet~$\Phi = (\phi, \psi_\alpha, F)$ of~$R$-charge~$\fr$ and central charge~$\fz$. The consistency of the transformation rules~\scomptrans\ implies that~$\Phi$ is embedded in a general multiplet~\sdef\ as follows:
\eqn\chiralgen{\eqalign{\Phi = \Big(& \phi, -  \sqrt 2 i \psi, 0, - 2 i F, 0, - i D_\mu \phi, (\fz - \fr H) \phi, \cr
& 0, 0, {\fr \over 4}  \big(R - 2 V^\mu V_\mu - 6 H^2 \big) \phi - \big(\fz - \fr H\big) H \phi \Big)~.}}
The supersymmetry transformations rules for the components of~$\Phi$ are then given by
\eqn\phitrans{\eqalign{&\delta \phi = \sqrt2 \zeta \psi~,\cr
& \delta \psi = \sqrt 2 \zeta F - \sqrt 2 i \left(\fz - \fr H\right) \t \zeta \phi - \sqrt 2 i \gamma^\mu \t \zeta \, D_\mu \phi~, \cr
& \delta F = \sqrt 2 i \left(\fz - (\fr - 2)H\right) \t \zeta \psi - \sqrt 2 i D_\mu \big(\t \zeta \gamma^\mu \psi\big)~.}}

\medskip

\item{2.)} {\it Anti-Chiral Multiplet:} The conjugate to~$\Phi$ is an anti-chiral multiplet~$\t \Phi = (\t \phi, \t \psi_\alpha, \t F)$ of~$R$-charge~$-\fr$ and central charge~$-\fz$, which is embedded in a general multiplet~\sdef\ with~$\chi_\alpha = 0$,
\eqn\achiralgen{\eqalign{\t \Phi = \Big(& \t \phi, 0, \sqrt 2i \t \psi, 0, 2 i \t F, i D_\mu \t \phi, \left(\fz - \fr H\right) \t \phi, \cr
& 0, 0, {\fr \over 4} \left(R - 2 V^\mu V_\mu - 6 H^2\right)\t \phi - \big(\fz - \fr H\big) H \t \phi \Big)~.}}
Its supersymmetry transformations are given by
\eqn\phitildetrans{\eqalign{& \delta \t \phi = - \sqrt 2 \t \zeta \t \psi~,\cr
& \delta \t \psi = \sqrt 2 \t \zeta \t F + \sqrt 2 i \left(\fz - \fr H\right) \zeta \t \phi + \sqrt 2 i \gamma^\mu \zeta \, D_\mu \t \phi~,\cr
& \delta \t F = \sqrt 2 i \left(\fz - (\fr -2 )H\right) \zeta \t \psi - \sqrt2 i D_\mu \big(\zeta \gamma^\mu \t \psi\big)~.}}

\medskip

\item{3.)} {\it Real Linear Multiplet:} If we set~$M = \t M = 0$ in~\sdef, we obtain a real linear multiplet~$\CJ = (J,   j_\alpha, \t j_\alpha, j_\mu, K)$. It has vanishing~$R$- and~$Z$-charge, $\fr = \fz = 0$, and~$j_\mu$ is a conserved current, so that~$\grad_\mu j^\mu = 0$. In terms of the general superfield~\sdef,
\eqn\lingen{\eqalign{\CJ = \Big( & J, j, \t j, 0, 0, - j_\mu - V_\mu J, -K, - {i \over 2} H \t j + i \gamma^\mu \big(D_\mu + {i \over 2}V_\mu\big) \t j,\cr
&  {i \over 2} H j - i \gamma^\mu \big(D_\mu - {i \over 2} V_\mu\big) j, - V^\mu j_\mu -  H K - \grad^\mu \grad_\mu J - V^\mu V_\mu J \Big)~.}}
The transformation rules are given by
\eqn\lintrans{\eqalign{& \delta J = i \zeta j + i \t \zeta \t j~,\cr
& \delta j = \t \zeta K + i  \gamma^\mu \t \zeta \big(j_\mu + i \d_\mu J + V_\mu J\big)~,\cr
& \delta \t j = \zeta K - i \gamma^\mu \zeta \big(j_\mu - i \d_\mu J +  V_\mu J\big)~,\cr
& \delta j_\mu = i \ep_{\mu\nu\rho} \grad^\nu \big(\zeta \gamma^\rho j - \t \zeta \gamma^\rho \t j\big)~,\cr
& \delta K = - i \grad_\mu \big(\zeta \gamma^\mu j + \t \zeta \gamma^\mu \t j\big) + 2 i H \big(\zeta j + \t \zeta \t j\big) - V_\mu \big(\zeta \gamma^\mu j - \t \zeta \gamma^\mu \t j\big)~.}}

\medskip

We can use the multiplets above to write general supersymmetric Lagrangians, i.e. scalar operators of vanishing~$R$- and~$Z$-charge that transform into a total derivative under supersymmetry transformations:

\medskip

\item{1.)} {\it D-Terms:} Consider a general superfield~$\CS$ with~$\fr = \fz = 0$. We see from~\scomptrans\ that the~$D$-component of~$\CS$ does not transform into a total derivative. However, it is straightforward to check that the supersymmetry variation of the following Lagrangian is a total derivative:
\eqn\ld{{\scr L}_D = -\half \left(D-a_\mu V^\mu - \sigma H \right)~.}
This is the curved-space analogue of a general~$D$-term Lagrangian.

\item{2.)} {\it F-Terms:} The~$F$-component of a chiral superfield~$\Phi$ with~$\fr = 2$ and~$\fz = 0$ has vanishing~$R$- and~$Z$-charge. Moreover, it follows from~\phitrans\ that the supersymmetry variation of~$F$ is a total derivative. Similarly, it follows from~\phitildetrans\ that the same is true for the~$\t F$-component of an anti-chiral superfield~$\t \Phi$ with~$\fr = -2$ and~$\fz = 0$. Therefore, we can write the curved-space analogue of a general~$F$-term Lagrangian,
\eqn\lf{{\scr L}_F = F + \t F~.}

\item{3.)} {\it Improvements of the~$R$-Current:} At the linearized level, we can perform an improvement~\imp\ of the~$\CR$-multiplet by a linear multiplet~$\CJ$.  This has the effect of coupling the flavor current~$j_\mu$ embedded in~$\CJ$ to the~$R$-symmetry gauge field~$A^\mu$. At the non-linear level, we can couple the conserved current~$j_\mu$ in the linear multiplet~\lingen\ to~$A^\mu$ and use~\lintrans\ to find suitable correction terms to obtain a supersymmetric Lagrangian:
\eqn\lj{{\scr L}_\CJ = j_\mu \Big(A^\mu - \half V^\mu\Big) + HK - {1 \over 4} \big(R - 2 V^\mu V_\mu + 2 H^2) J~.}

\medskip
\noindent Below, we will use these formulas to write supersymmetric Lagrangians for Abelian gauge theories coupled to chiral matter.

Finally, we will describe the multiplication rules for general multiplets. Given two such multiplets~$\CS_1, \CS_2$, whose bottom components~$C_1, C_2$ have~$R$-charges~$\fr_1, \fr_2$ and central charges~$\fz_1, \fz_2$, we would like to construct a multiplet~$\CS$ with bottom component~$C = C_1 C_2$. Using the transformation laws in~\scomptrans, we find that the components of~$\CS$ are given by
\eqn\sprod{\eqalign{& C = C_1 C_2~, \qquad \chi = \chi_1 C_2 + C_1 \chi_2~, \qquad \t \chi = \t \chi_1 C_2 + C_1 \t \chi_2~,\cr
& M = M_1 C_2 + C_1 M_2 - i \chi_1 \chi_2~, \qquad \t M = \t M_1 C_2 + C_1 \t M_2 - i \t \chi_1 \t \chi_2~,\cr
& { a}_\mu = { a}_{1 \mu} C_2 + C_1 { a}_{2 \mu} - \half \left(\chi_1 \gamma_\mu \t \chi_2 - \t \chi_1 \gamma_\mu \chi_2\right)~, \qquad \sigma = \sigma_1 C_2 + C_1 \sigma_2 + {i \over 2} \big(\chi_1 \t \chi_2 + \t \chi_1 \chi_2\big)~,\cr
& \lambda = \Big( \lambda_1 C_2  +{i \over 2} \t M_1 \chi_2 - \half \gamma^\mu \t \chi_1 \big({ a}_{2\mu} - i D_\mu  C_2\big) + {i \over 2} \t \chi_1 \left(\sigma_2 + \left(\fz_2 - \fr_2 H\right) C_2\right) \Big) + \left(1 \leftrightarrow 2\right)~,\cr
& \t \lambda = \Big( \t \lambda_1 C_2 - {i \over 2} M_1 \t \chi_2 - \half \gamma^\mu \chi_1 \big({ a}_{2\mu} + i D_\mu  C_2\big) - {i \over 2} \chi_1 \left(\sigma_2 - \left(\fz_2 - \fr_2 H\right)C_2\right)\Big) + \left(1 \leftrightarrow 2\right)~,\cr
& D = D_1 C_2 + C_1 D_2 + \half M_1 \t M_2 + \half \t M_1 M_2 - { a}^\mu_1 { a}_{2\mu} - \sigma_1 \sigma_2 - D^\mu C_1 D_\mu C_2 \cr
& \hskip20pt  + \left(\fz_1 - \fr_1 H \right)\left(\fz_2 - \fr_2 H\right) C_1C_2 - {i \over 2} \left(\left(\fz_1 - \fz_2\right) - \left(\fr_1 - \fr_2\right) H\right) \big(\chi_1 \t \chi_2 - \t \chi_1 \chi_2\big) \cr
& \hskip20pt - \chi_1 \big(\lambda_2 - {i \over 2} \gamma^\mu D_\mu \t \chi_2\big) + \t \chi_1 \big(\t \lambda_2 + {i \over 2} \gamma^\mu D_\mu \chi_2\big) - \big(\lambda_1 + {i \over 2} D_\mu \t \chi_1 \gamma^\mu\big) \chi_2  \cr
& \hskip20pt + \big(\t \lambda_1 - {i \over 2} D_\mu \chi_1 \gamma^\mu\big) \t \chi_2 - V^\mu \big(\chi_1 \gamma_\mu \t \chi_2 - \t \chi_1 \gamma_\mu \chi_2\big)~.}}

\subsec{Adding Gauge Fields}

In this subsection we explain how to accommodate gauge fields and charged chiral matter. For simplicity, we restrict ourselves to Abelian gauge fields. The extension to the non-Abelian case is straightforward.

A gauge multiplet~$\CV$ is a general multiplet~\sdef\ of vanishing~$R$- and~$Z$-charge, subject to the gauge freedom
\eqn\gaugeeq{\delta \CV = \Omega + \t \Omega~,}
where~$\Omega$ and~$\t \Omega$ are chiral and anti-chiral, respectively, and also have vanishing~$R$- and~$Z$-charge. We can use this gauge freedom to partially fix a Wess-Zumino gauge,
\eqn\vwzgauge{\CV = \left(0,0,0,0,0, a_\mu, \sigma, \lambda, \t \lambda, D\right)~.}
The residual gauge transformations that preserve this form of~$\CV$ are given by
\eqn\deltaa{\delta a_\mu = \d_\mu \Lambda^{(a)}~, \qquad \omega = - \t \omega = {i \over 2} \Lambda^{(a)}~,}
where~$\omega, \t \omega$ are the bottom components of~$\Omega, \t \Omega$. Therefore, $a_\mu$ is an Abelian gauge field.  We will denote its field strength by~$f_{\mu\nu} = \d_\mu a_\nu - \d_\nu a_\mu$. As before, the fact that we are working in Euclidean signature means that the gauge parameter~$\Lambda^{(a)}$ in~\deltaa\ is generally complex. In Wess-Zumino gauge, the supersymmetry transformations of~$\CV$ follow from~\scomptrans\ and~\vwzgauge,
\eqn\vtrans{\eqalign{& \delta a_\mu = - i \big(\zeta \gamma_\mu \t \lambda + \t \zeta \gamma_\mu \lambda\big)~,\cr
& \delta \sigma = - \zeta \t \lambda + \t \zeta \lambda~,\cr
& \delta \lambda = i \zeta \left( D + \sigma H\right) - {i\over 2} \ep^{\mu\nu\rho} \gamma_\rho \zeta \, f_{\mu\nu} - \gamma^\mu \zeta \left(i \d_\mu \sigma - V_\mu \sigma\right)~,\cr
&\delta \t \lambda = - i \t \zeta \left(D + \sigma H\right) - {i \over 2} \ep^{\mu\nu\rho} \gamma_\rho \t \zeta \, f_{\mu\nu} + \gamma^\mu \t \zeta \left(i \d_\mu \sigma + V_\mu \sigma\right)~,\cr
&\delta D = \grad_\mu \big( \zeta \gamma^\mu \t \lambda - \t \zeta \gamma^\mu \lambda \big) - i V_\mu \big(\zeta \gamma^\mu \t \lambda + \t \zeta \gamma^\mu \lambda\big) -  H \big(\zeta \t \lambda - \t \zeta \lambda\big)~.}}
As usual, the supersymmetry transformations~\scomptrans\ take us out of Wess-Zumino gauge, which must be restored by a compensating gauge transformation of the form~\gaugeeq. We will return to this point below.

As in flat space, it is convenient to define a gauge-invariant multiplet~$\Sigma$, whose bottom component is~$\sigma$. Using the supersymmetry variations in~\vtrans, we find that~$\Sigma$ is a real linear multiplet~\lingen\ with
\eqn\sigmacomp{\eqalign{& J = \sigma~, \qquad j = i \t \lambda~, \qquad \t j = - i \lambda~,\cr
& j_\mu = - {i\over 2} \ep_{\mu\nu\rho} f^{\nu\rho}~, \qquad K = D + \sigma H~.}}
Below, we will use~$\Sigma$ to write supersymmetric Yang-Mills and Chern-Simons terms.

Under a gauge transformation~\gaugeeq, a chiral multiplet~$\Phi$ of charge~$\fq$ and an anti-chiral multiplet~$\t \Phi$ of charge~$-\fq$ transform as follows:
\eqn\phigt{\delta \Phi = 2 \fq \Omega \Phi~, \qquad \delta \t \Phi = 2 \fq \t \Omega \t \Phi~.}
Here the chiral multiplets~$\Omega$ and~$\Phi$ are multiplied according to the rules in~\sprod, which gives another chiral multiplet, and similarly for~$\t \Omega, \t \Phi$. Note that under the residual gauge freedom~\deltaa\ that preserves Wess-Zumino gauge, we have the usual transformations~$\delta \Phi = i \fq \Lambda^{(a)} \Phi$ and~$\delta \t \Phi = - i \fq \Lambda^{(a)} \t\Phi$.

Since we will work in Wess-Zumino gauge throughout and this gauge is not preserved by the supersymmetry transformations derived in the previous subsection, we will need to accompany them by an appropriate gauge transformation to restore Wess-Zumino gauge. This modifies the supersymmetry transformation rules of charged fields. In particular, a chiral multiplet~$(\phi, \psi_\alpha, F)$ of charge~$\fq$, $R$-charge~$\fr$, and central charge~$\fz$ transforms as follows:
\eqn\chargedchiral{\eqalign{& \delta \phi = \sqrt 2 \zeta \psi~,\cr
& \delta \psi = \sqrt 2 \zeta F - \sqrt 2 i \left(\fz - \fq \sigma - \fr H\right) \t \zeta \phi - \sqrt 2 i \gamma^\mu \t \zeta D_\mu \phi~,\cr
& \delta F = \sqrt 2 i \left(\fz - \fq \sigma - (\fr - 2)H\right) \t \zeta \psi + 2 i\fq \phi  \t \zeta \t \lambda - \sqrt2 i D_\mu \big(\t \zeta \gamma^\mu \psi\big)~.}}
Here we have extended the definition of the covariant derivative~$D_\mu$ to also include the gauge field~$a_\mu$,
\eqn\hatdnew{D_\mu = \grad_\mu - i \fr \Big(A_\mu - \half V_\mu\Big) - i \fz C_\mu - i \fq a_\mu~.}
Therefore, the transformations~\chargedchiral\ are covariant with respect to all gauge redundancies. Note that~$\sigma$ only appears in the combination~$\fz - \fq \sigma$. We will return to the relation between~$\fz$ and~$\sigma$ below. The transformation rules for a charged anti-chiral multiplet are similarly modified.

Since charged fields have modified supersymmetry transformation rules, they also obey modified multiplication rules. In order to construct supersymmetric Lagrangians for charged fields, we will need to know how to multiply a chiral multiplet~$\Phi = (\phi, \psi_\alpha, F)$ of charge~$\fq$, $R$-charge~$\fr$, and central charge~$\fz$ with its conjugate~$\t \Phi = (\t \phi, \t \psi_\alpha, \t F)$.\foot{Alternatively, as in flat space, we could use the unmodified multiplication rules~\sprod\ and consider the gauge-invariant product multiplet~$\t \Phi e^{-2 \fq \CV} \Phi$, evaluated in Wess-Zumino gauge.} The relevant components of the product multiplet~$\CK = \t \Phi \Phi = (C^{(\CK)}, \chi^{(\CK)}_\alpha, \cdots)$, which does not carry any charge, are given by
\eqn\phiphit{\eqalign{& C^{(\CK)} = \t \phi \phi~,\cr
& a_\mu^{(\CK)} = - i\big(\t \phi D_\mu \phi - \phi D_\mu \t \phi\big) + \t \psi\gamma_\mu \psi~,\cr
&\sigma^{(\CK)} = 2\big(\fz - \fq \sigma - \fr H\big) \t \phi \phi + i \t \psi \psi~,\cr
& D^{(\CK)} = - 2 D^\mu \t \phi D_\mu \phi + i \t \psi \gamma^\mu D_\mu \psi - i D_\mu \t \psi \gamma^\mu \psi + 2 \t F F  \cr
& \hskip34pt - 2 \Big(\fq D + \big(\fz - \fr H\big) H + \left(\fz - \fq \sigma - \fr H\right)^2 - {\fr \over 4} \big(R - 2 V^\mu V_\mu - 6 H^2\big)\Big) \t \phi \phi \cr
& \hskip34pt- 2 i \Big(\fz - \fq \sigma - \fr H\Big) \t \psi \psi + 2 V^\mu \t \psi \gamma_\mu \psi - 2 \sqrt 2 i \fq \big(\t \phi \lambda \psi + \phi \t \lambda \t \psi\big)~.}}

So far we have discussed vector multiplets that contain dynamical gauge fields. We can also consider background vector multiplets, which couple to global flavor symmetries. While we are free to consider arbitrary classical configurations for these background vector multiplets, only certain configurations are consistent with rigid supersymmetry. By analogy with our treatment of background supergravity, these are the configurations for which the gauginos~$\lambda, \t \lambda$ as well as their supersymmetry variations~$\delta \lambda, \delta \t \lambda$ in~\vtrans\ vanish. (See~\FestucciaWS\ for a related discussion in four dimensions.) The only configurations compatible with four supercharges take the form
\eqn\gaugebg{a_\mu = - \sigma C_\mu + a_\mu^{(0)}~, \qquad \d_\mu \sigma = 0~, \qquad D = - \sigma H~,}
where~$a_\mu^{(0)}$ is a flat connection. (More general background gauge fields are allowed in the presence of fewer rigid supercharges.) As in flat space, real values of~$\sigma$ give rise to real masses for chiral superfields charged under~$\CV$ (see below). However, we are free to consider complex values of~$\sigma$ and~$a_\mu^{(0)}$ and this played an important role in~\refs{\FestucciaWS, \ClossetVG,\ClossetVP}.

\subsec{Supersymmetric Lagrangians}

In this section we use the formalism developed above to write down supersymmetric Lagrangians for gauge fields and chiral matter. We first consider supersymmetric Lagrangians for an Abelian gauge field~$\CV$ in Wess-Zumino gauge, as in~\vwzgauge. Starting with different multiplets constructed out of~$\CV$ and applying~\ld\ or~\lj, we obtain the following supersymmetric Lagrangians:
\medskip
\item{1.)} {\it Fayet-Iliopoulos Term:} Applying~\ld\ to~$-2 \xi \CV$, we obtain
\eqn\fiterm{{\scr L}_{ FI} =\xi \left(D -a_\mu V^\mu - \sigma H\right)~,}
where~$\xi$ is the dimension one Fayet-Iliopoulos parameter.

\item{2.)} {\it Gauge-Gauge Chern-Simons Term:} By applying~\ld\ to the multiplet~${k_{gg} \over 2 \pi} \CV \Sigma$, we obtain a level~$k_{gg}$ Chern-Simons term for the gauge field~$a_\mu$,
\eqn\lcs{{\scr L}_{gg} = {k_{gg} \over 4 \pi} \left(i \ep^{\mu\nu\rho} a_\mu \d_\nu a_\rho - 2 D \sigma + 2 i \t \lambda \lambda\right)~.}

\item{3.)} {\it Gauge-$R$ Chern-Simons Term:} By applying~\lj\ to the real linear multiplet~$\CJ = - {k_{gr} \over 2 \pi} \Sigma$, we obtain a mixed Chern-Simons term for the gauge field~$a_\mu$ and the background~$R$-symmetry gauge field~$A_\mu$,
    \eqn\frcs{{\scr L}_{gr} = {k_{gr} \over 2 \pi} \left(i \ep^{\mu\nu\rho} a_\mu \d_\nu \Big(A_\rho - \half V_\rho\Big) - DH + {1 \over 4} \sigma \big(R - 2 V^\mu V_\mu - 2 H^2\big)\right)~.}
    This term played an important role in~\refs{\ClossetVG, \ClossetVP}.

\item{4.)} {\it Yang-Mills Term:} If we apply~\ld\ to~$-{1 \over e^2} \Sigma^2$, where~$e$ is the gauge coupling, we obtain Yang-Mills kinetic terms for~$a_\mu$,
\eqn\YMi{\eqalign{{\scr L}_{YM} =~& {1\over 4e^2 }  f^{\mu\nu} f_{\mu\nu} + {1 \over 2 e^2} \partial^\mu \sigma\partial_\mu\sigma - {i\over e^2}\t\lambda \gamma^\mu\big(D_\mu + {i\over 2} V_\mu\big)\lambda \cr
& + {i \over 2 e^2}\sigma \ep^{\mu\nu\rho} V_\mu f_{\nu\rho} - {1 \over 2 e^2} V^\mu V_\mu \sigma^2 -{1 \over 2 e^2} \left(D+ \sigma H\right)^2 +{i\over 2 e^2} H\t\lambda \lambda~.}}

Finally, it is straightforward to obtain the Lagrangian for a chiral multiplet~$\Phi$ of charge~$\fq$, $R$-charge~$\fr$, and central charge~$\fz$, and its conjugate anti-chiral multiplet~$\t \Phi$, by applying~\ld\ to the product multiplet~$\CK = \t \Phi \Phi$ and using~\phiphit. Here we only consider the case where~$\Phi$ has canonical K\"ahler potential. Using the formalism developed above, it is straightforward to obtain the corresponding results for a general K\"ahler potential. The answer can be succinctly expressed using a modified covariant derivative,
\eqn\defDtilde{ {\scr D}_\mu = D_\mu + i  \fr_0 V_\mu = \grad_\mu - i\fr \Big(A_\mu-{3\over 2}V_\mu\Big) -i (\fr-\fr_0) V_\mu - i \fz C_\mu - i \fq a_\mu~.}
Here~$\fr_0$ is the superconformal~$R$-charge for a free chiral field, i.e.~$\fr_0(\phi) = \half$, $\fr_0 (\psi) = - \half$, and similarly for~$\t \phi, \t \psi$. The Lagrangian is given by
\eqn\philagfinal{\eqalign{{\scr L}_{\CK}  =~&  { \scr D}^\mu \t \phi {\scr  D}_\mu \phi - i \t \psi \gamma^\mu {\scr D}_\mu \psi - \t F F  + \fq (D + \sigma H) \t \phi \phi - 2 (\fr - 1) H (\fz - \fq \sigma) \t \phi \phi \cr
& + \Big( \left(\fz - \fq \sigma\right)^2  - {\fr \over 4} R + \half \big(\fr - \half\big ) V^\mu V_\mu  + \fr \big(\fr - \half\big) H^2 \Big) \t \phi \phi \cr
&+\Big(\fz - \fq \sigma - \big(\fr - \half\big) H\Big) i \t \psi \psi +  \sqrt 2 i \fq \big(\t \phi \lambda \psi + \phi \t \lambda \t \psi\big)~.}}
 When~$\phi$ has superconformal~$R$-charge~$r = r_0 = \half$ and central charge~$z =0$, the background fields~$A_\mu - \half V_\mu$ and~$H$, which are absent in conformal supergravity, drop out.

Note that if we view the gauge multiplet as a background field and turn on a configuration of the form~\gaugebg, then~$D= - \sigma H$ and the constant value of~$\sigma$ simply gives a contribution~$-\fq\sigma$ to the central charge~$\fz$. If~$\sigma = m$ is real and~$\fq = 1$, this corresponds to a real mass~$m$ for the chiral multiplet~$\Phi$. Thus, as in flat space, real masses for chiral multiplets correspond to suitable background gauge fields for flavor symmetries.

\newsec{Comparison with Linearized Supergravity}

In this section we expand the background fields discussed in section~4 and the supersymmetric Lagrangians in section~6 around flat space. In this limit, we can also derive them using the linearized supergravity formulas of section 2 and our knowledge of the~$\CR$-multiplet in flat space. This comparison serves as a non-trivial check of our results and elucidates their structure in terms of familiar flat-space objects. We also comment on the fact that the first-order deformation of the Lagrangian around flat space is~$Q$-exact.

\subsec{Linearized Background Fields}

In an expansion around flat space, $g_{\mu\nu} = \delta_{\mu\nu} + 2 h_{\mu\nu}$, the linearized couplings to new minimal supergravity are given by~\rhcomp,
\eqn\rhcompbis{{\delta \scr L} = - T_{\mu\nu} h^{\mu\nu}+ j_\mu^{(R)} \big(A^\mu - {3 \over 2} V^\mu \big) + j_\mu^{(Z)} C^\mu + J^{(Z)} H~.}
Here we have omitted terms containing the gravitini, which are set to zero in the rigid limit. We would like to determine the background fields~$A_\mu, V_\mu, H$ that are needed to preserve a given supercharge~$\zeta$. At leading order around flat space, we are free to choose
\eqn\zetaconst{\zeta_\alpha = \pmatrix{1 \cr 0}~,}
which corresponds to~$s = 1$ in~\zetasol. It suffices to find linear combinations of~$T_{\mu\nu}$ and the other bosonic operators in the~$\CR$-multiplet that are invariant under the flat-space supercharge~$\delta_Q$ corresponding to~$\zeta$.

Using~\rmultcomp, we  find that the components of the~$\CR$-multiplet transform as follows:
\eqn\dqrmult{\eqalign{& \delta_Q j_\mu^{( R )} = - i \zeta S_\mu~,\cr
& \delta_Q S_\mu = 0~,\cr
& \delta_Q \t S_\mu = \zeta \big(2 j_\mu^{(Z)} + i \ep_{\mu\nu\rho} \d^\nu j^{( R ) \rho} \big)+ \gamma^\nu \zeta \big(2 i T_{\mu\nu} + \d_\nu j_\mu^{( R )} - \ep_{\mu\nu\rho} \d^\rho J^{(Z)} \big)~,\cr
& \delta_Q T_{\mu\nu} = {i \over 4} \ep_{\mu\rho\lambda} \zeta \gamma^\rho \d^\lambda S_\nu + {i \over 4} \ep_{\nu\rho\lambda} \zeta \gamma^\rho \d^\lambda S_\mu~,\cr
& \delta_Q j_\mu^{(Z)} = -{i \over 2} \zeta \gamma^\nu \d_\nu S_\mu - \half \ep_{\mu\nu\rho} \zeta \d^\nu S^\rho~,\cr
& \delta_Q J^{(Z)} = - \half \zeta \gamma^\mu S_\mu~.}}
As in section~4, we switch from the usual flat-space coordinates~$x^1, x^2, x^3$ to the adapted coordinates\foot{Here we define~$z, \b z$ so that the usual orientation~$\ep_{123} =1$ corresponds to~$\ep^{\tau z \b z} = 2i$.}
\eqn\flatcoord{\tau = x^1~, \qquad z = x^2 - i x^3~, \qquad \b z = x^2 + i x^3~,}
so that the almost contact structure corresponding to~$\zeta$ in~\zetaconst\ is given by~$\eta = \d_\tau$. Using the transformation laws in~\dqrmult, we find that the following operators are invariant under~$\delta_Q$, i.e. they are~$Q$-closed:
\eqn\qclosed{\eqalign{& T_{\tau \tau} - i j_\tau^{(Z)} + 2 i \d_z j_{\b z}^{(R)}~,\cr
& T_{\tau z} - i j_z^{(Z)} - {i \over 2} \d_z j_\tau^{(R)} - \half \d_z J^{(Z)}~,\cr
& T_{\tau \b z} - i j_{\b z}^{(Z)} - i \d_\tau j_{\b z}^{(R)} + {i \over 2} \d_{\b z} j_{\tau}^{(R)} + \half \d_{\b z} J^{(Z)}~,\cr
& T_{zz} - {i \over 2} \d_z j_z^{(R)}~,\cr
& T_{z \b z} - {i \over 2} \d_z j_{\b z}^{(R)} - {1 \over 4} \d_\tau J^{(Z)}~,\cr
& j_z^{(Z)} - i \d_z J^{(Z)}~.}}
There is, however, no~$Q$-invariant combination that contains~$T_{\b z \b z}$. The operators in~\qclosed\ are also~$Q$-exact. They are proportional to the six operators~$\delta_Q \t S_{\mu\alpha}$ in~\dqrmult. We will return to this point below.

Using the operators in~\qclosed\ we find that the following Lagrangian is~$Q$-invariant:
\eqn\lqinv{\eqalign{\delta {\scr L} =~&  - 2 h^{\tau z} \big(T_{\tau z} - i j_z^{(Z)} - {i \over 2} \d_z j_\tau^{(R)} - \half \d_z J^{(Z)}\big) \cr
& - 2 h^{\tau \b z} \big(T_{\tau \b z} - i j_{\b z}^{(Z)} - i \d_\tau j_{\b z}^{(R)} + {i \over 2} \d_{\b z} j_{\tau}^{(R)} + \half \d_{\b z} J^{(Z)}\big)\cr
&- 2 h^{z\b z} \big(T_{z \b z} - {i \over 2} \d_z j_{\b z}^{(R)} - {1 \over 4} \d_\tau J^{(Z)}\big) + f^z \big( j_z^{(Z)} - i \d_z J^{(Z)}\big)~.}}
Here~$f^z(\tau, z, \b z)$ is a complex function and we have set~$h_{\tau\tau} = 0$ by a gauge transformation in order to compare with the formulas in section~4. The absence of a term proportional to~$h^{\b z \b z}$ is due to the fact that there is no~$Q$-invariant operator containing~$T_{\b z \b z}$. Since the metric is real, this also means that~$h^{zz}$ is absent. Integrating by parts and comparing with~\rhcompbis, we can determine the background fields~$A_\mu, V_\mu, H$ in terms of~$f^z$ and the linearized metric,
\eqn\linbgs{\eqalign{& V^\tau = 2 i \big(\d_z h^{\tau  z} - \d_{\b z} h^{\tau \b z} \big) + \d_z f^z~, \cr
&  V^z = - 2 i \d_\tau h^{\tau z} -\d_\tau f^z~, \qquad V^{\b z} = 2 i \d_\tau h^{\tau \b z}~,\cr
& H = \d_{\b z} h^{\tau \b z} - \d_z h^{\tau z} - \half \d_\tau h^{z\b z} + i \d_z f^z~,\cr
& A_\tau = 4 i \big(\d_z h_{\tau \b z} - \d_{\b z} h_{\tau z} \big) + {3 \over 2} \d_z f^z~,\cr
& A_z = i \d_\tau h_{\tau z} -2 i \d_z h_{z \b z}~, \qquad A_{\b z} = - 3 i \d_\tau h_{\tau \b z} - {3 \over 4}\d_\tau f^z~.}}
The metric-dependent terms precisely agree with the linearization of~\auxsol, \difcob, and~\Asol. The function~$f^z$ corresponds to the linearized shifts in~\shiftbck\ with~$U^z=-\d_\tau f^z$ and~$\kappa= \d_z f^z$. Note that the condition~\lucond\ is automatically satisfied. (See appendix~B of~\DumitrescuHA\ for a related discussion in four dimensions.)

The fact that the linear terms in the background fields are~$Q$-exact does not guarantee that the same is true at the non-linear level, where both the Lagrangian and the supersymmetry transformations are modified. Understanding these non-linear terms is essential in order to establish the parameter dependence of the partition function on supersymmetric manifolds. We will present this analysis elsewhere.

\subsec{Linearized Lagrangians}

In the previous subsection we used the general structure of the~$\CR$-multiplet to reproduce the supersymmetric background fields of section~4 at the linearized level. Similarly, we can reproduce the supersymmetric Lagrangians of section~6 at the linearized level by examining the~$\CR$-multiplet for gauge multiplets and chiral matter.

\medskip

\item{1.)} {\it Gauge Multiplet:} Let~$\CV$ be a~$U(1)$ vector superfield and consider the Lagrangian
\eqn\flatgaugel{{\scr L} = \int d^4 \theta \, \left(-{1 \over e^2} \Sigma^2 + {k \over 2 \pi} \CV \Sigma - 2 \xi \CV\right)~,}
where~$\Sigma = {i \over 2} \t D D \CV$ is the field-strength superfield, $k$ is the Chern-Simons level, and~$\xi$ is the Fayet-Iliopoulos parameter. This Lagrangian is the flat-space limit of~\fiterm, \lcs, and~\YMi. It leads to the equation of motion~${i\over 2 e^2} \t D D \Sigma = {k \over 2 \pi} \Sigma - \xi$, which gives the following~$\CR$-multiplet:
\eqn\rmultym{\CR_{\alpha\beta} = {2 \over e^2} \big(D_\alpha \Sigma \t D_\beta \Sigma + D_\beta \Sigma \t D_\alpha \Sigma\big)~, \qquad \CJ^{(Z)} = - \xi \Sigma + {i \over 4 e^2} \t D D \big(\Sigma^2\big)~.}
Note that the Chern-Simons term does not contribute to the~$\CR$-multiplet. We will need the following bosonic components:
\eqn\ymrmult{\eqalign{& j_\mu^{(R)} = -{1 \over e^2} \t \lambda \gamma_\mu \lambda~, \cr
&   j_\mu^{(Z)}   = {1 \over 2 e^2} \d^\nu \big(2 \sigma  f_{\mu\nu}-i\ep_{\mu\nu\rho}\, \t \lambda \gamma^\rho \lambda\big) +{i\,\xi\over 2} \, \ep_{\mu\nu\rho} f^{\nu\rho}~, \cr
&  J^{(Z)} = - \xi \sigma - {1 \over e^2} \big(\sigma D - {i \over 2} \t \lambda \lambda\big)~.}}
Substituting into~\rhcompbis, we find perfect agreement with the first-order terms in~\fiterm\ and~\YMi. The Chern-Simons term~\lcs\ does not contain any first-order terms, consistent with the fact that it does not contribute to the~$\CR$-multiplet. In this example we have not discussed the energy-momentum tensor~$T_{\mu\nu}$, since it can be obtained by the usual procedure of minimally coupling the flat-space Lagrangian~\flatgaugel\ to the metric. We will now discuss an example where this is no longer the case.

\medskip

\item{2.)} {\it Chiral Multiplet:} Let~$\Phi$ be a chiral superfield of~$R$-charge~$\fr$ and real mass~$m$, and let~$\t \Phi$ be its conjugate anti-chiral superfield. Consider the Lagrangian
    \eqn\flatchil{{\scr L} =  \int d^4 \theta \, \t \Phi e^{- 2i m \t \theta \theta} \Phi~,}
    which is the flat-space limit of~\philagfinal, with~$\fz = -m$ and~$\fq = 0$. The equations of motion~$D^2 \big(e^{- 2i m \t \theta \theta} \Phi\big) = \t D^2 \big(e^{- 2i m \t \theta \theta} \t \Phi\big) = 0$ imply that the~$\CR$-multiplet is given by
    \eqn\rmultfchi{\eqalign{& \CR_{\alpha\beta} = \Big(e^{2im\t \theta \theta} D_\alpha \big(e^{-2i m \t \theta \theta} \Phi\big) \t D_\beta \big(e^{-2im \t \theta \theta} \t \Phi\big) - {\fr \over 2} [D_\alpha, \t D_\beta] \CJ \Big) + \left(\alpha \leftrightarrow \beta\right)~,\cr
    & \CJ^{(Z)} = - m \CJ - {i \over 2} \Big(\fr - \half \Big)  \t D D\CJ~.}}
    Here~$\CJ = \t \Phi e^{- 2i m \t \theta \theta} \Phi$ is a real linear multiplet, $D^2 \CJ = \t D^2 \CJ = 0$, which contains the conserved current~$j_\mu = i \big (\t \phi \d_\mu \phi - \d_\mu \t \phi \phi\big) - \t \psi \gamma_\mu \psi$ associated with the global~$U(1)$ flavor symmetry of~\flatchil. Note that the terms in~\rmultfchi\ proportional to the~$R$-charge~$\fr$ take the form of an improvement transformation~\imp\ by~$\CJ$. The bosonic components of the~$\CR$-multiplet~\rmultfchi\ are given by
    \eqn\chiboscomp{\eqalign{& j_\mu^{(R)} = i \fr \big(\t \phi \d_\mu \phi - \d_\mu \t \phi \phi\big) - (\fr - 1) \t \psi \gamma_\mu \psi~,\cr
    & T_{\mu\nu} = \cdots + {\fr \over 2} \big(\d^2 \delta_{\mu\nu} - \d_\mu \d_\nu\big) \t \phi \phi~,\cr
    & J^{(Z)} = - m \t \phi \phi + \Big(\fr - \half\Big) \big(2 m \t \phi \phi - i \t \psi \psi \big)~,\cr
    & j_\mu^{(Z)} = - m j_\mu - i \Big(\fr - \half\Big) \ep_{\mu\nu\rho} \d^\nu j^\rho~.}}
Here the ellipsis denotes the usual, unimproved energy-momentum tensor that follows by minimally coupling the flat-space Lagrangian~\flatchil\ to a background metric. As above, we can substitute these operators into~\rhcompbis\ and find perfect agreement with the first-order terms in~\philagfinal\ with~$\fz = -m$ and~$\fq = 0$. In particular, the improvement term in~$T_{\mu\nu}$ reproduces the coupling to the Ricci scalar~$R$.

\newsec{The Energy-Momentum Tensor Two-Point Function}

In this section we will show how the flat-space two-point functions of the~$R$-current and the energy-momentum tensor in~$\CN=2$ superconformal theories can be computed using localization on the squashed three-sphere~$S_b^3$ discussed in section~5.2. Recall that in four-dimensional~$\CN=1$ SCFTs, the two-point functions of currents are related to anomalies, and hence calculable. For instance, the two-point function of the energy-momentum tensor is determined by the~$c$-anomaly. Even though there are no local anomalies in three dimensions, these two-point functions are nevertheless calculable using localization.

\subsec{Two-Point Functions from Squashing}

Given an~$\CN=2$ SCFT, we can place the theory on the squashed three-sphere~$S^3_b$ by coupling its superconformal~$\CR$-multiplet to the background fields in~\metsqst. Recall that the~$\CR$-multiplet is not unique in the presence of Abelian flavor symmetries. The superconformal~$\CR$-multiplet can be determined using~$F$-maximization on the round three-sphere~\refs{\JafferisUN,\JafferisZI,\ClossetVG}, which also allows us to extract the two-point functions of the Abelian flavor currents in the SCFT~\ClossetVG. Here we will consider the SCFT on the squashed sphere~$S^3_b$ and extract the two-point functions of the~$R$-current and the energy-momentum tensor.

We will study the partition function~$Z_{S^3_b}$ and the associated free energy,
\eqn\freeen{F(b)=-\log Z_{S^3_b}~,}
as a function of the squashing parameter~$b$ near~$b=1$, which corresponds to the round~$S^3$. If the SCFT arises in the deep IR of an RG flow from a free theory in the UV, then~$F(b)$ can be computed exactly using localization~\refs{\HamaEA,\ImamuraWG}. We will show that
\eqn\tauRRtob{\Re {\del^2 F\over \del b^2} \bigg|_{b=1}={\pi^2\over 2}\tau_{rr}~,}
where the coefficient~$\tau_{rr}>0$ determines the two-point functions of the~$R$-current and the energy-momentum tensor in flat space and at separated points (see for instance~\ClossetVP),
\eqn\twopointRR{\eqalign{& \langle j_\mu^{(R)} (x) j_\nu^{(R)}(0)\rangle = {\tau_{rr}\over 16\pi^2}\left(\delta_{\mu\nu}\del^2-\del_\mu\del_\nu\right) {1\over x^2}~,\cr
&\langle T_{\mu\nu} (x) T_{\rho\sigma}(0)\rangle =  -{\tau_{rr}\over 64\pi^2}
(\delta_{\mu\nu}\d^2-\d_\mu\d_\nu)(\delta_{\rho\sigma}\d^2-\d_\rho\d_\sigma){1\over x^2}  \cr
&\hskip77pt +{\tau_{rr}\over 64\pi^2} \left((\delta_{\mu\rho}\d^2-\d_\mu\d_\rho)((\delta_{\nu\sigma}\d^2-\d_\nu\d_\sigma)+(\mu\leftrightarrow\nu)\right){1\over x^2}~.}}
The fact that both correlators are given in terms of~$\tau_{rr}$ follows from superconformal invariance. A free chiral multiplet has~$\tau_{rr} = {1 \over 4}$.

It is convenient  to use coordinates~$x^\mu$ on~$S^3_b$ that reduce to stereographic coordinates on~$S^3$ when~$b=1$.\foot{In this section, we raise and lower the indices on the coordinates~$x^\mu$ using~$\delta_{\mu\nu}$, so that~$x^\mu x_\mu = \delta_{\mu\nu} x^\mu x^\nu$. Tensor indices are raised and lowered using the curved metric $g_{\mu\nu}$, as usual.} The metric \metsqst\ on the squashed sphere then takes the form
\eqn\metcf{\eqalign{&g_{\mu\nu}=\Omega^2 \, \delta_{\mu\nu}+\left({b-b^{-1}\over b+b^{-1}}\right)^2v_\mu v_\nu~, \qquad \Omega={2  \over 1+x^\mu x_\mu}~,\cr
&  v_\mu dx^\mu=\Omega^2 \, {b+b^{-1}\over 2 }\left( (x^1dx^2- x^2 dx^1)+(1- \Omega^{-1})dx^3 + x^3 x_\mu  dx^\mu \right)~,}}
where we have set the radius~$r=1$ in~\metsqst. The vector $v^\mu$ is Killing and satisfies~$v^\mu v_\mu=1$. In these coordinates, the other background fields are given by
\eqn\aux{A_\mu=V_\mu= {\big(b-b^{-1}\big)} v_\mu~,\qquad C_\mu=i \left({b-b^{-1}\over b+b^{-1}}\right)v_\mu~,\qquad H={ i(b+b^{-1})\over 2 }~.}
If the squashed sphere is nearly round, so that~$b = 1 + \delta b$ with~$|\delta b | \ll 1$, then the background fields~$A_\mu$ and~$V_\mu$ are of order~$\delta b$, while~$g_{\mu\nu}$ and~$H$ differ from their values on the round~$S^3$ by terms of order~$(\delta b)^2$.

We will use the conformal mapping of primary operators from flat space to the round~$S^3$ to understand small deformations of the theory on~$S^3_b$ around~$b=1$. At order~$\delta b$, the Lagrangian is modified by the operators that couple to~$A_\mu$ and $V_\mu$,
\eqn\vcouplestojR{\delta {\scr L}_1 = \big(A^\mu-{3\over 2}V^\mu\big) j_\mu^{(R)}+j_\mu^{(Z)} C^\mu= \delta b \,  v^\mu\big( i j_\mu^{(Z)} - j_\mu^{(R)}\big)~.}
At order~$(\delta b)^2$, the Lagrangian is modified by some operator~$\CO$, whose one-point function on the round~$S^3$ must vanish by conformal invariance,
\eqn\ordertwo{\delta {\scr L}_2= (\delta b)^2 \CO~, \qquad \langle \CO\rangle_{S^3}=0~.}
Using~\vcouplestojR\ and~\ordertwo, we find that
\eqn\relderFandviv{ {\del^2 F\over \del b^2} \bigg|_{b=1} = - \int_{S^3} d^3x \sqrt{g}  \int_{S^3} d^3y \sqrt{g} \; v_\mu(x)v_\nu(y) \langle  j^{( R )\mu}(x)  j^{( R ) \nu}(y)\rangle_{S^3} + \big({\rm contact~ terms}\big)~.}
The first derivative of~$F(b)$ is determined by a one-point function, and hence it vanishes at~$b=1$ by conformal invariance. In deriving~\relderFandviv, we have omitted the contribution of the operator~$j_\mu^{(Z)}$ in~\vcouplestojR, since it is redundant in the SCFT, i.e.~its correlation functions vanish at separated points. However, both~$j_\mu^{(Z)}$ and~$j_\mu^{( R )}$ may give rise to contact terms. As we will explain below, these can only affect the imaginary part~$\Im F(b)$. To study the second derivative of the real part~$\Re F(b)$ we only need the two-point function of~$j_\mu^{( R )}$ at separated points.

Using a conformal transformation, we can map the flat-space two-point function of~$j_\mu^{( R )}$ in~\twopointRR\ to the round~$S^3$. At separated points, we find
\eqn\twopointjjc{\langle  j^{( R )\mu}(x)  j^{( R )\nu}(y)\rangle_{S^3} = {\tau_{rr}\over 4\pi^2}  \, {1\over s(x,y)^6} \, \left(\delta^{\mu\nu} (x-y)^2 -2(x-y)^\mu (x-y)^\nu \right)~,}
where~$s(x,y)$ is the $SO(4)$ invariant distance function on~$S^3$,
\eqn\defsxy{
s(x,y) = {2 |x-y|\over \sqrt{1+x^2} \sqrt{1+y^2}}~.
}
Substituting into~\relderFandviv, we find that the integral is UV divergent and must be regulated.  The answer is finite and unambiguous due to current conservation,\foot{See section~3 of~\ClossetVG, where a similar issue is discussed in detail.} and it reduces to~\tauRRtob.

To complete the argument, we must analyze the effects of possible contact terms in~\relderFandviv. All contact terms in correlation functions of operators in the~$\CR$-multiplet are captured by local terms in the background supergravity fields, which may contribute to~$F(b)$. These local terms were classified in~\refs{\ClossetVP\ClossetVG}. Their reality properties are fixed by the corresponding contact terms in flat space. On the squashed sphere, they only contribute to the imaginary part~$\Im F(b)$, and hence they cannot affect~\tauRRtob.

\subsec{Examples}

We will illustrate~\tauRRtob\ in two examples: a free chiral multiplet and any large-$N$ superconformal theory with an~$AdS_4\times X^7$ supergravity dual.

The partition function~$F_{\Phi}(b)$ for a free massless chiral multiplet~$\Phi$ of superconformal~$R$-charge~$\fr=\ha$ was computed in~\ImamuraWG\ using localization. A useful representation of~$F_{\Phi}(b)$ is given by (see for instance~\refs{\VdBthesis,\BeniniMF})
\eqn\intrepFPhi{
F_{\Phi}(b) = - \int_0^\infty {d x\over 2x} \left( {\sinh x \over \sinh (b x)\sinh (  b^{-1}x )} -  {1 \over x}  \right)~,
}
so that
\eqn\Fddbff{
{\d^2 F_{\Phi} \over \d b^2}\bigg|_{ b=1}= {\pi^2\over 8}~.}
This agrees with~\tauRRtob\ when~$\tau_{rr}={1\over 4}$, which is the correct value for a free chiral multiplet.

Consider an~$\CN=2$ Chern-Simons-matter theory with an~$AdS_4\times X^7$ supergravity dual. (Here~$X^7$ is a Sasaki-Einstein manifold.) In the large-$N$ limit, the dependence on the squashing parameter simplifies dramatically~\ImamuraWG,
\eqn\FblargeN{F(b)= {(b+b^{-1})^2\over 4} \; F(1)~,}
while the free energy~$F(1)$ on the round~$S^3$ is simply related to~$\tau_{rr}$~\BarnesBW,
\eqn\ftaurel{F(1)= {\pi^2\over 4}\tau_{rr}~,}
Again, we find perfect agreement with~\tauRRtob.

\vskip 1cm

\noindent {\bf Acknowledgments:}
We would like to thank Francesco Benini, Stefano Cremonesi, Yakov Eliashberg, Daniel Jafferis, Dario Martelli, Martin Rocek, Adam Schwimmer, Nathan Seiberg, Itamar Shamir, James Sparks, and especially Maxim Kontsevich for useful discussions. CC is a Feinberg postdoctoral fellow at the Weizmann Institute of Science and would like to thank the Institute for Advanced Study for its hospitality during the completion of this project. The work of TD was supported in part by a DOE Fellowship in High Energy Theory and a Centennial Fellowship from Princeton University. The work of GF was supported in part by NSF grant PHY-0969448 and a Marvin L. Goldberger Membership at the Institute for Advanced Study.  ZK was supported by NSF grant PHY-0969448, a research grant from Peter and Patricia Gruber Awards, a grant from the Robert Rees Fund for Applied Research, as well as by the Israel Science Foundation under grant number~884/11.  ZK  would also like to thank the United States-Israel Binational Science Foundation (BSF) for support under grant number~2010/629. GF and ZK are grateful for the hospitality of the Aspen Center for Physics during the completion of
this project. Any opinions, findings, and conclusions or recommendations expressed in this
material are those of the authors and do not necessarily reflect the views of the funding agencies.

\appendix{A}{Conventions}

\subsec{Flat Euclidean Space}

The metric is given by~$\delta_{\mu\nu}$, with $\mu,\nu=1,2,3$. The totally antisymmetric Levi-Civita symbol is normalized so that~$\ep_{123}=1$. A spinor~$\psi_\alpha$ transforms as a doublet of~${\rm Spin}(3) =  SU(2)$. Spinor indices are raised and lowered by acting on the left with the antisymmetric symbols~$\ep^{\alpha\beta}, \ep_{\alpha\beta}$, which are normalized so that~$\ep^{12}= \ep_{21} = 1$. When spinor indices are omitted, all spinor products are to be read as follows:
\eqn\spcon{\psi\chi=\psi^\alpha \chi_\alpha~.}
The Hermitian conjugate spinor~$\psi^\dagger$, which also transforms as a doublet of $SU(2)$, is defined with the following index structure:
\eqn\daggers{\big(\psi^\dagger\big)^\alpha = \overline {(\psi_\alpha)}~, }
where the bar denotes complex conjugation. On anti-commuting spinors, we take Hermitian conjugation to be order reversing. We use~$\psi, \t \psi$ to denote spinors that would be conjugate in Lorentzian signature. In Euclidean signature, they are independent.

The gamma matrices are given by
\eqn\choicegam{
{(\gamma^\mu)_{\alpha}}^{\beta}=(\sigma^3, -\sigma^1, -\sigma^2)~,
}
where~$\sigma^1, \sigma^2, \sigma^3$ are the Pauli matrices. They satisfy the following identities:
\eqn\gmmattd{\eqalign{
& \gamma^\mu\gamma^\nu = \delta^{\mu\nu} + i \ep^{\mu\nu\rho}\gamma_\rho~,\cr
& \big(\gamma^\mu\big)_{\alpha\beta} \big(\gamma_\mu\big)_{\kappa \lambda} = - \big(\ep_{\alpha\kappa} \ep_{\beta\lambda} + \ep_{\alpha\lambda} \ep_{\beta \kappa}\big)~.
}}

The~$\CN=2$ anti-commuting superspace coordinates are denoted by~$\theta_\alpha, \t \theta_\alpha$, and the supercovariant derivatives are given by
\eqn\ddtdef{D_\alpha = {\d \over \d \theta^\alpha} - i (\gamma^\mu \t \theta)_\alpha \d_\mu~, \qquad \t D_\alpha = - {\d \over \d \t \theta^\alpha} + i (\gamma^\mu \theta)_\alpha \d_\mu~.}
We also define the superspace integrals
\eqn\inttheta{\int d^2\theta \, \theta^2=1~,\qquad \int d^2 {\t \theta} \,  \t \theta^2= 1~, \qquad \int d^4 \theta \, \theta^2 \t \theta^2 = 1~.}

\subsec{Differential Geometry}

Lowercase Greek letters~$\mu,\nu,\ldots$ denote curved indices, while lowercase Latin letters~$a,b,\ldots$ denote frame indices. Given a Riemannian metric~$g_{\mu\nu}$, we can define an orthonormal frame~$e^a_\mu$,
\eqn\mettetr{g_{\mu\nu}=\delta_{a b} e^a_\mu e^b_\nu~,\qquad g^{\mu\nu} e^a_\mu e^b_\nu=\delta^{ab}~.}
We will also denote~$g = |\det (g_{\mu\nu})|$. The spin connection corresponding to the Levi-Civita connection~$\grad_\mu$ is given by
\eqn\defspincon{
{\omega_{\mu a}}^b = e^b_\nu \nabla_\mu e_a^\nu~.
}
The Riemann tensor takes the form
\eqn\riemann{{R_{\mu\nu a}}^b = \d_\mu {\omega_{\nu a}}^b - \d_\nu {\omega_{\mu a}}^b + {\omega_{\nu a}}^c {\omega_{\mu c}}^b - {\omega_{\mu a}}^c {\omega_{\nu c}}^b~.}
The Ricci tensor is defined by~$R_{\mu\nu} = {R_{\mu\rho\nu}}^\rho$, and~$R = {R_\mu}^\mu$ is the Ricci scalar. In these conventions, the Ricci scalar is negative on a round sphere, i.e. $R=-6$ for the round unit sphere. In three dimensions the Weyl tensor vanishes, so that
\eqn\riethreed{R_{\mu\nu\rho\lambda}=R_{\mu\rho} g_{\nu\lambda}-R_{\mu\lambda} g_{\nu\rho} -R_{\nu\rho} g_{\mu\lambda} +R_{\nu\lambda} g_{\mu\rho} -\half R \left(g_{\mu\rho}  g_{\nu\lambda} - g_{\mu\lambda} g_{\nu\rho}\right)~.}
The covariant derivative of a spinor~$\psi$ is given by
\eqn\covspin{\nabla_\mu \psi = \big(\d_\mu - {i\over 4}\omega_{\mu ab} \ep^{abc}\gamma_c\big)\psi~,}
so that
\eqn\commcov{\ep_{\mu\nu\rho}[\grad^\nu,\grad^\rho]\psi={i\over 2}(2 R_{\mu\nu}-R g_{\mu\nu})\gamma^\nu \psi~.}
The Lie derivative of~$\psi$ along a vector~$X= X^\mu\d_\mu$ is given by~\Kosmann,
\eqn\Lieder{
 \CL_X \psi =X^\mu \grad_\mu  \psi +{i\over 4} (\grad_\mu X_\nu) \ep^{\mu\nu\rho}\gamma_\rho \psi~.
}
On tensors, the Lie derivative is defined as usual. For instance, the Lie derivative of~${\Phi^\mu}_\nu$ along~$\eta^\mu$ in~\liecondd\ is given by
\eqn\lied{\CL_\eta {\Phi^\mu}_\nu=\eta^\rho \grad_\rho {\Phi^\mu}_\nu-  \grad_\rho \eta^\mu {\Phi^\rho}_\nu+ \grad_\nu \eta^\rho  {\Phi^\mu}_\rho~.}

\appendix{B}{Almost Contact Structures}

In this appendix we review some properties of almost contact (metric) structures in three dimensions. (For additional background, see for instance~\Blair.) We also introduce a family of connections that are compatible with a given almost contact metric structure~(ACMS). Finally, we study the consequences of the integrability condition~\liecondd.

\subsec{Basic Definitions}

Given a nowhere vanishing one-form~$\eta_\mu$, a vector field~$\xi^\mu$, and a~$(1,1)$ tensor~${\Phi^\mu}_\nu$ on an orientable three-manifold~$\CM$, the triple~$(\eta, \xi, \Phi)$ defines an almost contact structure, if the following conditions are satisfied:
\eqn\concont{\eta_\mu\xi^\mu=1~,\qquad   {\Phi^\mu}_\rho {\Phi^\rho}_\nu=-{\delta^\mu}_\nu+\xi^\mu \eta_\nu~.}
The endomorphism~$\Phi$ has rank two, and its left and right kernels are generated by~$\eta$ and~$\xi$, respectively: if~$X^\mu$ satisfies~${\Phi^\mu}_\nu X^\nu=0$, then it is proportional to~$\xi^\mu$, and if~$\Omega_\mu$ satisfies~$\Omega_\mu {\Phi^\mu}_\nu=0$, then it is proportional to~$\eta_\mu$.

The vectors orthogonal to~$\eta_\mu$ define a sub-bundle~$\CD$ of the tangent bundle. Then~$\Phi$ induces an almost complex structure~$J = \Phi |_{\CD}$ on $\CD$, i.e.~$J^2 = \Phi^2 |_{\CD} = - \1$. (In this sense, an almost contact structure is the three-dimensional analogue of an almost complex structure.) We can use~$J$ to split~$\CD$ into holomorphic and anti-holomorphic vectors. This can also be expressed directly in terms of~$\Phi$: a vector~$X^\mu$ is holomorphic if and only if~${\Phi^\mu}_\nu X^\nu=i X^\nu$. Similarly, a one-form~$\Omega_\mu$ is holomorphic if and only if~$\Omega_\mu {\Phi^\mu}_\nu= i \Omega_\nu$.

If the manifold~$\CM$ is endowed with a Riemannian metric~$g_{\mu\nu}$, we say that the almost contact structure~$(\eta, \xi, \Phi)$ is compatible with~$g_{\mu\nu}$ if the following conditions hold:
\eqn\compatibility{\xi^\mu=\eta^\mu~,\qquad g_{\rho\lambda} {\Phi^\rho}_\mu {\Phi^\lambda}_\nu = g_{\mu\nu}-\eta_\mu \eta_\nu~.}
This defines an almost contact metric structure (ACMS) on~$\CM$. The existence of such a structure is equivalent to a reduction of the structure group of the tangent bundle to~$U(1)$. On an orientable three-manifold, an ACMS is equivalent to a choice of metric and a nowhere vanishing one-form~$\eta_\mu$. In this case, there is a choice of orientation such that~$\xi^\mu = \eta^\mu$ and~${\Phi^\mu}_\nu={\ep^\mu}_{\nu\rho} \eta^\rho$ satisfy~\concont\ and~\compatibility. This shows that every orientable Riemannian three-manifold admits a metric-compatible almost contact structure.

\subsec{Compatible Connections}

Given an ACMS, it is convenient to introduce a compatible connection~$\hat \grad_\mu$~\Sasaki. In three dimensions, it is sufficient to impose~$\hat \grad_\mu g_{\nu\rho} = 0$ and~$\hat \grad_\mu \eta_\nu = 0$. In general, a connection~$\hat \grad_\mu$ is metric compatible if and only if its connection coefficients can be expressed in terms of~$g_{\mu\nu}$ and the contorsion tensor~${K^\mu}_{\nu\rho}$,
\eqn\hatdef{{\, \hat \Gamma^\mu}_{\nu\rho} = \half g^{\mu\lambda} \left(\d_\nu g_{\rho\lambda} + \d_\rho g_{\nu\lambda} - \d_\lambda g_{\nu\rho}\right) + {K^\mu}_{\nu\rho}~, \qquad K_{\mu\nu\rho} = -K_{\rho\nu\mu}~.}
When the contorsion vanishes, we recover the Levi-Civita connection~$\grad_\mu$. The spin connection corresponding to~$\hat \grad_\mu$ is given by
\eqn\Sascon{\hat \omega_{\mu\nu\rho}= \omega_{\mu\nu\rho}-K_{\nu\mu\rho}~.}

The most general metric-compatible connection that also satisfies~$\hat \grad_\mu \eta_\nu = 0$ has contorsion
\eqn\crst{K_{\mu\nu\rho}= \eta_\mu \grad_\nu \eta_\rho-\eta_\rho \grad_\nu \eta_\mu-2 W_\nu \Phi_{\mu\rho}~,}
where~$W_\nu$ is any smooth one-form. Since this connection preserves the ACMS defined by~$\eta_\mu$, its holonomy is contained in~$U(1)$. When~$\eta_\mu$ is given in terms of a spinor~$\zeta$ as in~\etadef, the covariant derivative~$\hat \grad_\mu \zeta$ is given by
\eqn\covzeta{\hat\grad_{\mu}\zeta= \big(\grad_\mu-{i} W_\mu\big) \zeta-\ha (\grad_\mu \eta_\nu) \gamma^\nu \zeta~.}
Here we have not assumed that $\zeta$ satisfies the Killing spinor equation~\ksei. In section~4 we choose the one-form~$W_\mu$ as follows:
\eqn\defS{W_\mu=-{1\over 4}\eta_\mu \ep^{\nu\rho\lambda}\eta_\nu \d_\rho \eta_\lambda~.}

\subsec{Integrability Condition}

Here we study the integrability condition~\liecondd,
\eqn\liecondbis{{\Phi^\mu}_\nu \big( \CL_\xi {\Phi^\nu}_\rho \big) = 0~.}
Even though we arrived at this condition via~\etacond, which requires a metric, the condition~\liecondbis\ only depends on the almost contact structure~$(\eta, \xi, \Phi)$. Given a holomorphic one-form~$\Omega_\mu$, such that~$\Omega_\mu {\Phi^\mu}_\nu = i \Omega_\nu$, we can take the Lie derivative along~$\xi$ and use~\liecondbis\ to obtain
\eqn\omegacond{\left(\CL_\xi \Omega_\mu \right) {\Phi^\mu}_\nu = i \CL_\xi \Omega_\nu~.}
This shows that~$\CL_\xi \Omega_\mu$ is also a holomorphic one-form.

We will now show that there are local coordinates~$(\tau, z, \b z)$, which are adapted to the almost contact structure, so that~$\xi = \d_\tau$ and~$\Omega = \Omega_{z} d z$. The existence of such adapted coordinates constitutes a three-dimensional analogue of~\Newlander. Since~$\xi$ is nowhere vanishing, we can always choose coordinates~$x^1 = \tau, x^2, x^3$ such that~$\xi = \d_\tau$. Moreover,~$\xi^\mu \Omega_\mu = 0$, so that~$\Omega = \Omega_2 dx^2 + \Omega_3 dx^3$. We define the complex ratio function~$\rho= {\Omega_2 \over \Omega_3}$, which is smooth, non-vanishing, and does not depend on the choice of~$\Omega_\mu$, i.e. it is determined in terms of~${\Phi^\mu}_\nu$. Therefore, the holomorphic one-forms~$\Omega_\mu$ and~$\CL_\xi \Omega_\mu$ have identical ratio functions, which implies that~$\d_\tau \rho = 0$. It follows from classical theorems in the theory of linear partial differential equations (see for example~\Nirenberg, chapter~I and \Courant,  chapter 4, \S 8) that there exist complex coordinates~$z (x^2, x^3)$ such that~$\d_2 -\rho(x^2,x^3) \d_3 =\lambda(z,\b z) \partial_z $ for some non-vanishing $\lambda(z,\b z)$. Hence, there is a function~$\Omega_{z}(\tau, z , \b z)$, such that~$\Omega = \Omega_{z} d{ z}$.

We have thus shown that~\liecondbis\ implies the existence of an adapted cover for~$\CM$, with charts~$(\tau, z ,\b z)$, such that~$\xi = \d_\tau$ and holomorphic one-forms are proportional to~$d z$. Two overlapping adapted charts~$(\tau, z, \b z)$ and~$(\tau', z', \b z')$ are related by~$\tau'=\tau+t(z,\b z)$ and~$z'=f(z)$, where~$t(z, \b z)$ is real and~$f(z)$ is holomorphic.

Conversely, if~$\CM$ can be covered by adapted charts, it admits an ACMS satisfying the integrability condition~\liecondbis. In~$(\tau, z, \b z)$ coordinates, we have\foot{Here we choose the orientation on~$\CM$ such that~$\ep^{\tau z \b z}=i g^{-\ha}$.}
 \eqn\phicomp{{\Phi^z}_z=-{\Phi^{\b z}}_{\b z}=i~,\qquad {\Phi^{\b z}}_{ z}={\Phi^{ z}}_{\b z}=0~,}
from which~\liecondbis\ follows. Moreover, we can use adapted coordinates to rewrite~\liecondbis\ in terms of the almost complex structure~$J$ induced on~$\CD$ in the following intuitive way:
 \eqn\DJ{\d_\tau J=0~.}

So far our discussion of the integrability condition~\liecondbis\ has not involved the metric. If~$(\eta, \xi,\Phi)$ are part of an ACMS, the compatible metric in adapted coordinates takes the following form:
\eqn\acmsmets{ds^2=\big(d\tau+ h(\tau,z,\b z) dz+\b h(\tau,z,\b z) d \b z\big)^2+c(\tau,z,\b z)^2 dz d\b z~.}

We conclude with several comments:
\medskip
\item{1.)} Given any almost contact structure~$(\eta,\xi, \Phi)$, we can find coordinates~$x^1 = \tau, x^2, x^3$ such that~$\xi = \d_\tau$. However, a compatible metric cannot always be put into the form~\acmsmets. Instead, the most general such metric takes the form
\eqn\genmet{ds^2 = \big(d \tau + h_i dx^i\big)^2 + g^{(2)}_{ij} \, dx^i dx^j~,}
where~$i, j = 2, 3$ and~$h_i, g^{(2)}_{ij}$ are functions of~$\tau, x^2, x^3$. The integrability condition~\liecondbis\ implies the existence of a complex function~$z(x^2, x^3)$, such that
\eqn\mettrans{g^{(2)}_{ij}(\tau, x^2, x^3) dx^i dx^j = c(\tau, z, \b z)^2 dz d\b z~.}
\medskip
\item{2.)} Given any metric, we can choose local coordinates such that
\eqn\localmet{ds^2 = \big(dx^1 + h_i dx^i\big)^2 + c(x^1,x^2,x^3)^2 \big((dx^2)^2 + (dx^3)^2\big)~,}
but in general~$\xi \neq \d_1$. Locally, we can define a new, compatible almost contact structure~$\xi '= \d_1$, which satisfies the integrability condition~\liecondbis. Therefore, this condition does not restrict the metric locally. However, there may be global obstructions that prevent us from extending~$\xi'$ to a globally well-defined almost contact structure on~$\CM$. This should be contrasted with the four-dimensional case discussed in~\DumitrescuHA, where the metric is locally restricted by the condition that~$\CM$ be a Hermitian manifold.
\medskip
\item{3.)} We would like to point out the relation of~\liecondbis\ to the standard condition of normality for almost contact structures (see for instance \Blair). In three dimensions, normality is equivalent to~$\CL_\xi {\Phi^\mu}_\nu=0$, which is considerably stronger than~\liecondbis.

\appendix{C}{The Rigid Limit of Four-Dimensional New Minimal Supergravity}

Here we briefly summarize the key formulas describing four-dimensional~$\CN=1$ theories with a~$U(1)_R$ symmetry in curved superspace~\FestucciaWS, i.e. the rigid limit of new minimal supergravity in four-dimensions~\refs{\SohniusTP,\SohniusFW}. We follow the conventions of~\DumitrescuHA, except that we use uppercase Latin letters~$M, N, \ldots$ to denote four-dimensional curved indices.

In four dimensions, the bosonic fields in the new minimal supergravity multiplet are the metric~$G_{MN}$, the~$R$-symmetry gauge field~$A_M$, and a covariantly conserved vector~$V^M$. The Killing spinor equations that govern the rigid limit are given by
\eqn\fdspeq{\eqalign{&(\grad_M - i A_M) \zeta = - i V_M \zeta - i V^N \sigma_{MN} \zeta~,\cr
& (\grad_M + iA_M)\t \zeta = i V_M \t \zeta + i V^N \t \sigma_{MN} \t \zeta~.}}
These equations were thoroughly studied in~\refs{\KlareGN,\DumitrescuHA}, where it was shown that a solution exists on any Hermitian manifold.

The transformation rules for a general multiplet~$(C, \chi_\alpha, \t \chi^\alphadot, M, \t M, \CA_M, \lambda_\alpha, \t \lambda^\alphadot, D)$ of~$R$-charge~$\fr$ take the following form:
\eqn\fdgmul{\eqalign{& \delta C = i \zeta \chi - i \t \zeta \t \chi~,\cr
& \delta \chi = \zeta M + \sigma^M \t \zeta \, (i\CA_M + D_M C)~,\cr
& \delta \t \chi = \t \zeta \t M + \t \sigma^M \zeta\, (i \CA_M - D_M C)~,\cr
& \delta M = 2 \t \zeta \t \lambda + 2 i D_M (\t \zeta \, \t \sigma^M \chi)~,\cr
& \delta \t M = 2 \zeta \lambda + 2 i D_M (\zeta \sigma^M \t \chi)~,\cr
& \delta \CA_M = i (\zeta \sigma_M \t \lambda + \t \zeta\, \t \sigma_M \lambda) + D_M (\zeta \chi + \t \zeta \t \chi)~,\cr
& \delta \lambda = i \zeta D + 2 \sigma^{MN}\zeta \, D_M \CA_N~, \cr
& \delta \t \lambda = - i \t \zeta D + 2 \t \sigma^{MN} \t \zeta \, D_M \CA_N~,\cr
& \delta D = - D_M (\zeta \sigma^M \t \lambda - \t \zeta \, \t \sigma^M \lambda) + 2 i V_M (\zeta \sigma^M \t \lambda + \t \zeta \, \t \sigma^M \lambda) \cr
& \hskip25pt + {i \fr \over 4} (R - 6 V^M V_M) (\zeta \chi + \t \zeta \t \chi)~,}}
where~$D_M  = \d_M - i \fr A_M$. The supersymmetry algebra is given by
\eqn\fdsalg{\eqalign{& \{\delta_\zeta, \delta_{\t \zeta}\} \varphi_\fr = 2 i \CL_K^{(A)} \varphi_\fr~, \qquad K^M = \zeta \sigma^M \t \zeta~,\cr
& \{\delta_\zeta, \delta_\eta\}\varphi_\fr = 0~, \qquad \{\delta_{\t \zeta}, \delta_{\t \eta}\} \varphi_\fr = 0~,}}
where~$\varphi_\fr$ is any field of~$R$-charge~$\fr$ and~$\CL^{(A)}_K = \CL_K - i \fr K^M A_M$ the~$R$-covariant Lie derivative. In order to verify that the transformations~\fdgmul\ satisfy the algebra~\fdsalg, we must use the fact that the spinor parameters satisfy the Killing spinor equations~\fdspeq.

\appendix{D}{Twisted Reduction to Three Dimensions}

In this appendix, we formally obtain any three-dimensional supersymmetric background, as well as the supersymmetry algebra and superfield transformation laws, by a twisted reduction from four dimensions.

\subsec{Review of Twisted Dimensional Reduction}

Here we closely follow~\ScherkZR. We use upper- and lowercase symbols for four- and three-dimensional quantities, respectively. When there is potential for confusion, we designate a three-dimensional quantity using a superscript~`$(3)$'. Curved and frame indices are denoted by~$M, N, \ldots$,~$\mu, \nu, \ldots$ and~$ A,  B, \ldots$,~$ a,  b, \ldots$, respectively.
Consider four-dimensional coordinates
\eqn\fdcoord{X^M = (x^1, y, x^2, x^3)~.}
We will perform the reduction along the~$y$-coordinate. In our conventions,
\eqn\redsigma{ \sigma^a_{\alpha \dot \beta} = \gamma^a_{\alpha\beta}~, \qquad  \sigma^2_{\alpha \dot \beta} = i \ep_{\alpha\beta}~,\qquad (\t \sigma^a)^{\dot \beta \alpha} =(\gamma^a)^{\beta \alpha}~, \qquad   (\t \sigma^2)^{\dot \beta \alpha }  = i \ep^{\beta\alpha}~.}

Consider a four-dimensional metric with Killing vector~$Y = \d_y$,
\eqn\fdmet{dS^2 = G_{MN}(x) dX^M dX^N =  \big(dy + c_\mu (x) dx^\mu\big)^2 + g_{\mu\nu}(x) dx^\mu dx^\nu~,}
which describes a non-trivial fibration of the~$y$-coordinate over a three-manifold with metric~$ds^2 = g_{\mu\nu} dx^\mu dx^\nu$. The inverse metric is given by
\eqn\invmet{G^{\mu\nu} = g^{\mu\nu}~, \qquad G^{y\mu} = - g^{\mu\nu} c_\nu~, \qquad G^{yy} = 1 + g^{\mu\nu} c_\mu c_\nu~,}
where~$g^{\mu\nu}$ is the inverse of~$g_{\mu\nu}$. After choosing a three-dimensional frame~$e^{  a}_\mu$ with inverse~$e^\mu_{  a}$, we define the four-dimensional frame~$E^{ A}_M$ and its inverse~$E_{  A}^M$,
\eqn\fdframe{E^{  A}_M = \pmatrix{e^{  a}_\mu & c_\mu\cr 0 & 1}~, \qquad E_{  A}^M = \pmatrix{e^\mu_{  a} & - e^\mu_{  a} c_\mu\cr 0 & 1}~.}
Here the last row and column of~$E^A_M$ correspond to~$M = y$ and~$A=2$, respectively. Note that~$\det(E_M^{  A}) = \det (e^{  a}_\mu)$, so that~$\det(G_{MN}) = \det(g_{\mu\nu})$. The spin connections~$\Omega_{M   A   B}$ and~$\omega_{\mu   a   b}$ are then related as follows:
\eqn\spinconnred{\eqalign{& \Omega_{\mu  a  b} = \omega_{\mu  a  b} + \half c_\mu e^\nu_{  a}e^{\rho}_{  b} (\d_\nu c_\rho - \d_\rho c_\nu)~,\cr
& \Omega_{\mu   2   b} = \half e^\nu_{  b} (\d_\mu c_\nu - \d_\nu c_\mu)~,\cr
&\Omega_{y   a   b} = \half e^\nu_{  a} e^\rho_{  b} (\d_\nu c_\rho - \d_\rho c_\nu)~,\cr
& \Omega_{y   2   b} = 0~.}}

The form of the metric~\fdmet\ is invariant under infinitesimal four-dimensional diffeomorphisms generated by a vector field~$\Xi^M (x)$ that is invariant along~$Y$, i.e.~$[Y,\Xi] = 0$. We distinguish between three-dimensional diffeomorphisms parametrized by~$\Xi^\mu(x)$, under which~$g_{\mu\nu}$, $c_\mu$, $e^{  a}_\mu$, and~$e_{  a}^\mu$ transform as three-tensors, and diffeomorphisms parametrized by~$\Lambda^{( c)}(x) = \Xi^y(x)$, under which~$g_{\mu\nu}$, $e^{  a}_\mu$, and~$e_{  a}^\mu$ are invariant, but~$c_\mu$ transforms as a gauge field,
\eqn\hgaugetrans{\delta c_\mu = \d_\mu \Lambda^{( c)}(x)~.}
This identifies~$c_\mu$ as the graviphoton.

Given a four-dimensional vector~$U^M(x)$ that is invariant along~$Y$, so that~$[Y, U] = 0$, we can define a three-dimensional vector~$u^\mu(x) = U^\mu(x)$ and scalar~$u = U^y + U^\mu c_\mu$. Similarly, given a four-dimensional one-form~$W_M(x)$ that satisfies~$\CL_Y W = 0$, we can construct a three-dimensional one-form~$w_\mu = W_\mu - W_y c_\mu$  and scalar~$w = W_y$.
These rules apply to all tensor indices. For gauge fields, we take the gauge parameters to be  invariant along~$Y$.

Consider a~$y$-dependent four-dimensional scalar field~$\Phi(x, y)$,
\eqn\fdcc{\Phi(x, y) = e^{i \fz y} \phi(x)~.}
This means that~$\Phi$ has central charge~$\fz$ under graviphoton gauge transformations~\hgaugetrans. Upon reduction to three dimensions, the derivative~$W_M = \d_M \Phi$ automatically becomes gauge covariant, so that~$w_\mu = (\d_\mu - i \fz c_\mu) \Phi$ and~$w = i \fz \Phi$. We will also need the derivatives of spinors~$\chi_\alpha$, $\t \chi^\alphadot$ with central charge~$\fz$,
\eqn\morederivs{\eqalign{& \grad_\mu \chi = \grad_\mu^{(3)} \chi + {1 \over 4} \ep_{\mu\nu\rho} v^\nu \gamma^\rho \chi + {1 \over 4} c_\mu v^\nu \gamma_\nu \chi~, \qquad\quad \grad_y \chi = i z \chi + {1 \over 4} v^\mu \gamma_\mu \chi~,\cr
& \grad_\mu \t \chi  = \grad_\mu^{(3)} \t \chi - {1 \over 4} \ep_{\mu\nu\rho} v^\nu \gamma^\rho \t \chi + {1 \over 4} c_\mu v^\nu \gamma_\nu \t \chi~, \qquad\quad \grad_y \t \chi = i z \t \chi + {1 \over 4} v^\mu \gamma_\mu \t \chi~.}}
Here~$\grad^{(3)}_\mu$ is the Levi-Civita connection corresponding to the three-dimensional metric~$g_{\mu\nu}$ and~$v^\mu$ is the dual graviphoton field strength,
\eqn\gpfsdef{v^\mu = -i \ep^{\mu\nu\rho} \d_\nu c_\rho~, \qquad \grad^{(3)}_\mu v^\mu = 0~.}
Finally, we will need the four-dimensional Ricci scalar~$R$ in terms of~$R^{(3)}$ and~$v^\mu$,
\eqn\riccired{R = R^{(3)} - \half v^\mu v_\mu~.}

\subsec{Reduction of the Killing Spinor Equation}

Starting with the rigid limit of four-dimensional new minimal supergravity reviewed in appendix~C, we re-derive the three-dimensional Killing spinor equations~\ksei\ and~\kseii\ (see also~\KlareGN).

Consider a four-dimensional spinor~$\zeta$ that satisfies the first equation in~\fdspeq,
\eqn\fdspeqbis{(\grad_M  - i A_M) \zeta = - i V_M - i V^N \sigma_{MN} \zeta~.}
Since~$A_M, V_M$, and~$\zeta$ do not carry central charge, we assume that they are invariant along~$Y$, so that~$\grad_\mu^{(3)} V^\mu = 0$ and~$\zeta(x)$ does not depend on~$y$ in the frame~\fdframe.

We can now use~\morederivs\ to obtain the~$M = y$ component of~\fdspeqbis,
\eqn\ycomp{{1 \over 4} v^\mu \gamma_\mu \zeta = i (A_y - V_y) \zeta + \half V^\mu \gamma_\mu \zeta~,}
where~$v^\mu$ is the dual graviphoton field strength defined in~\gpfsdef. In order to satisfy this equation without imposing additional restrictions on~$\zeta$, we set
\eqn\moretddefs{H = A_y = V_y~, \qquad V^\mu = \half v^\mu~.}
The~$M = \mu$ component of~\fdspeqbis\ then takes the form
\eqn\mucomp{\grad^{(3)}_\mu \zeta = i(a_\mu - v_\mu) \zeta - \half H \gamma_\mu \zeta - \half \ep_{\mu\nu\rho} v^\nu \gamma^\rho \zeta~,}
where we have defined the three-dimensional~$R$-symmetry gauge field~$a_\mu$ via
\eqn\tdrsymmgf{A^\mu = a^\mu - \half v^\mu~.}
With these definitions, the equation~\mucomp\ is identical to the Killing spinor equation~\ksei.

It is straightforward to carry out the reduction for a four-dimensional spinor~$\t \zeta$ that satisfies the second equation in~\fdspeq. This leads to the same rules~\moretddefs\ and~\tdrsymmgf\ for the reduction of the supergravity background fields. Therefore, any supersymmetric background in four dimensions formally leads to a three-dimensional background that preserves the same number of supercharges. However, as we saw in section~4,  the three-dimensional graviphoton that is needed to solve~\mucomp\ is generally complex, while the graviphoton obtained from the four-dimensional metric is necessarily real. Therefore, we must carry out the reduction before complexifying the graviphoton.

\subsec{Reduction of the Supersymmetry Transformations}

We start with the transformation rules~\fdgmul\ for a four-dimensional multiplet of~$R$-charge~$\fr$. We assign central charge~$\fz$ to all fields in the multiplet by taking their~$y$-dependence to be~$e^{i \fz y}$. If we define~${\bf a}^\mu = \CA^\mu$,~$\sigma = \CA_y$,~$D^{(3)} = D - \sigma H$, we obtain the following three-dimensional supersymmetry transformation rules:
\eqn\tdgmulapp{\eqalign{& \delta C = i \zeta \chi + i \t \zeta \t \chi~,\cr
& \delta \chi = \zeta M - \t \zeta \left(\sigma + \left(\fz - \fr H\right) C\right) - \gamma^\mu \t \zeta \big(D^{(3)}_\mu C + i {\bf a}_\mu\big)~,\cr
& \delta \t \chi = \t \zeta \t M - \zeta \left(\sigma - \left(\fz - \fr H\right) C\right) - \gamma^\mu \zeta \big(D^{(3)}_\mu C -i {\bf a}_\mu \big)~,\cr
& \delta M = - 2 \t \zeta \t \lambda + 2 i \left(\fz - \left(\fr - 2\right) H\right) \t \zeta \chi - 2 i D^{(3)}_\mu \big(\t \zeta \gamma^\mu \chi\big)~,\cr
& \delta \t M = 2 \zeta \lambda - 2 i \left(\fz - \left(\fr + 2\right) H\right) \zeta \t \chi - 2 i D^{(3)}_\mu \big(\zeta \gamma^\mu \t \chi\big)~,\cr
& \delta {\bf a}_\mu = - i \big(\zeta \gamma_\mu \t \lambda + \t \zeta \gamma_\mu \lambda\big) + D_\mu^{(3)} \big(\zeta \chi - \t \zeta \t \chi\big)~,\cr
& \delta \sigma = - \zeta \t \lambda + \t \zeta \lambda + i \left(\fz - \fr H\right) \big(\zeta \chi - \t \zeta \t \chi\big)~,\cr
& \delta \lambda = i \zeta \big(D^{(3)} + \sigma H\big) - i \ep^{\mu\nu\rho} \gamma_\rho \zeta \, D^{(3)}_\mu {\bf  a}_\nu - \gamma^\mu \zeta \big((\fz - \fr H) {\bf a}_\mu + i D_\mu^{(3)} \sigma - v_\mu \sigma\big)~,\cr
&\delta \t \lambda = - i \t \zeta \big(D^{(3)} + \sigma H\big) - i \ep^{\mu\nu\rho} \gamma_\rho \t \zeta \, D_\mu^{(3)} {\bf a}_\nu + \gamma^\mu \t \zeta \big((\fz - \fr H) {\bf a}_\mu + i D_\mu^{(3)} \sigma + v_\mu \sigma\big)~,\cr
&\delta D^{(3)} = D_\mu^{(3)} \big( \zeta \gamma^\mu \t \lambda - \t \zeta \gamma^\mu \lambda \big) - i v_\mu \big(\zeta \gamma^\mu \t \lambda + \t \zeta \gamma^\mu \lambda\big)  \cr
& \hskip36 pt - H \big(\zeta \t \lambda - \t \zeta \lambda\big) + \left(\fz - \fr H\right) \Big(\zeta \t \lambda + \t \zeta \lambda- i H \big(\zeta \chi - \t \zeta \t \chi\big)\Big) \cr
& \hskip36pt + {i \fr \over 4} \big(R^{(3)} - 2v^\mu v_\mu - 6 H^2\big)\big(\zeta \chi - \t \zeta \t \chi\big)~,}}
where
\eqn\dthree{D^{(3)}_\mu  = \grad^{(3)}_\mu - i \fr \Big(a_\mu - \half v_\mu\Big) - i \fz c_\mu~.}
We have thus recovered the transformation rules~\scomptrans.

Similarly, the algebra satisfied by the transformation rules~\tdgmulapp\ is the dimensional reduction of the four-dimensional algebra~\fdsalg,
\eqn\tdsalgapp{\eqalign{& \{ \delta_\zeta, \delta_{\t \zeta}\} \varphi_{(\fr,\fz)} = - 2i \big(\CL'_{K} \varphi_{(\fr,\fz)}  +  \zeta \t \zeta \left(\fz - \fr H\right) \varphi_{(\fr,\fz)} \big)~, \qquad K^\mu = \zeta \gamma^\mu \t \zeta~,\cr
& \{ \delta_\zeta, \delta_\eta\} \varphi_{(\fr,\fz)} = 0~, \qquad \{ \delta_{\t \zeta}, \delta_{\t \eta}\} \varphi_{(\fr,\fz)}= 0~,}}
in perfect agreement with~\rigidsalg. Here~$\CL_K'$ is the twisted Lie derivative defined in~\hatliedef.

\listrefs

\end